\documentclass[journal]{IEEEtran}

\usepackage{amssymb}
\usepackage{diagbox}
\usepackage{cite, setspace,graphicx,graphics, subfigure, amsmath,amssymb,amsfonts,stfloats,multirow,booktabs, url}
\usepackage[below]{placeins}
\DeclareGraphicsExtensions{.pdf,.jpeg,.png}
\ifCLASSINFOpdf
\else
\fi

\ifCLASSOPTIONcompsoc
 \usepackage[caption=false,font=normalsize,labelfont=sf,textfont=sf]{subfig}
\else
 \usepackage[caption=false,font=footnotesize]{subfig}
\fi

\begin{document}
\title{Reversible Data Hiding in JPEG Images with Multi-objective Optimization}
\author{Zhaoxia~Yin,~\IEEEmembership{Member,~IEEE},~Yuan Ji,~Bin Luo
\thanks{This research work is partly supported by National Natural Science Foundation of China (61872003, U1636206,61860206004).}
\thanks{Zhaoxia Yin, Yuan Ji and Bin Luo are with the school of Computer Science and Technology, Anhui University, e-mail: yinzhaoxia@ahu.edu.cn, luobin@ahu.edu.cn.}
\thanks{Corresponding author: Bin Luo.}
}


\maketitle

\begin{abstract}
Among various methods of reversible data hiding (RDH) in JPEG images, the consideration in designing is only the image quality, but the image quality and the file size expansion are equally important in JPEG images.
Based on this situation, we propose a RDH scheme in JPEG images considering both the image quality and the file size expansion while designing the algorithm. The multi-objective optimization strategy is utilized to realize the balance of the two objectives. Specifically, the cover is divided into several non-overlapping signals firstly, and after that, the embedding costs of signals are calculated using the knowledge of the JPEG compression. Next, the optimized combination of signals for embedding data is gained by the multi-objective optimization. Experimental results show the better performance of our proposed RDH compared with state-of-the-art RDH in JPEG images.
\end{abstract}

\IEEEpeerreviewmaketitle

\section{Introduction}\label{Introduction}
\IEEEPARstart{N}{owadays}, data is abundant and exists everywhere, and the importance of data privacy is getting more and more attention, so the technology of data hiding is developing fast. The traditional way of protecting data is encryption, but it can expose the transmission way of additional data. Then, data hiding appeared to make up for this shortcoming, it can hide the additional data into cover medium which can be shared publicly, so the behavior of transferring additional data is invisible. As technology advances, some applications have more requirements for data hiding, they may need both the additional data and the cover be lossless. That is to say, the cover must be fully recovered after the additional data is extracted. This technology is named reversible data hiding (RDH) \cite{IEEEexample:ni2003reversible,IEEEexample:1227616,IEEEexample:lin2008multilevel,IEEEexample:sachnev2009reversible,IEEEexample:tsai2009reversible,IEEEexample:chen2013reversible,IEEEexample:li2015efficient,IEEEexample:wang2017rate,IEEEexample:8283771}, it can be employed in many fields, such as military, medicine, forensics and so on.
\par Image is a frequently used form of data in our daily life, so many RDH algorithms are designed on images. Furthermore, the Joint Photographic Experts Group (JPEG) is a widely used image format in the network, thus the research of RDH in JPEG images is very popular. The RDH in spatial domain is growing rapidly, but the corresponding algorithms cannot be directly utilized in JPEG images, that is because the RDH in spatial domain is designed using the redundancy of images and the JPEG images are obtained by compressing the redundancy of spatial domain images which means there is less redundant space in JPEG images. In addition, the RDH in spatial images needn't consider the file size, but the RDH in JPEG images which are compressed format must take it into consideration. Here, table \ref{tb1differences of RDH} gives the differences of RDH in spatial domain and JPEG domain. Obviously, the file size is also an important metric for RDH in JPEG images except for the image quality and the payload because the compression of images is to reduce the file size of images.
\begin{table}
\caption{The differences of RDH in spatial domain and JPEG domain.}
\label{tb1differences of RDH}
\resizebox{250pt}{2.1cm}{
\begin{tabular}{|c|l|l|c|}
\hline
                                                                   & \multicolumn{1}{c|}{\begin{tabular}[c]{@{}c@{}}Operation \\ object\end{tabular}}  & \multicolumn{1}{c|}{\begin{tabular}[c]{@{}c@{}}Optimized\\  goal\end{tabular}}                           & \begin{tabular}[c]{@{}c@{}}Redundant \\ space\end{tabular} \\ \hline
\begin{tabular}[c]{@{}c@{}}RDH in\\  spatial\\ domain\end{tabular} & 1.pixels                                                                          & \begin{tabular}[c]{@{}l@{}}1.rate and \\ distortion\\ performance\end{tabular}                           & more                                                       \\ \hline
\begin{tabular}[c]{@{}c@{}}RDH in \\ JPEG\\ domian\end{tabular}    & \begin{tabular}[c]{@{}l@{}}1.DCT \\ coefficients\\ 2.entropy \\ code\end{tabular} & \begin{tabular}[c]{@{}l@{}}1.rate and \\ distortion\\ performance\\ 2.file size\\ expansion\end{tabular} & less                                                       \\ \hline
\end{tabular}
}
\end{table}
\par RDH in JPEG images has also developed in recent years. It can be divided into four categories. The first one is based on the lossless compression first proposed in \cite{IEEEexample:fridrich2002lossless}, this paper includes not only method based on lossless compression but also the data hiding approach through modifying quantization table according to their parity. The next method is based on the quantization table modification \cite{IEEEexample:fridrich2002lossless,IEEEexample:wang2013high}. Wang ${ et~al. }$ proposed a RDH scheme that modifies the DCT coefficients to embed data while the corresponding values in the quantization table make changes \cite{IEEEexample:wang2013high}. It can achieve good effects both in capacity and image quality, but the file size is relatively large. The third method is based on modifying the Huffman table \cite{IEEEexample:qian2012lossless,IEEEexample:hu2013improved,IEEEexample:du2018improved} which can well keep the file size of JPEG image unchanged. However, its embedding capacity is rare. The fourth category is the method based on the modification of quantized DCT coefficients \cite{IEEEexample:huang2016reversible,IEEEexample:wedaj2017improved,IEEEexample:hong2018improved,IEEEexample:xie2018reversible,IEEEexample:hou2018reversible,IEEEexample:liu2018reversible,IEEEexample:xuan2019minimum}. In 2016, Huang ${ et~al. }$ \cite{IEEEexample:huang2016reversible} applied the histogram shifting (HS) in the RDH scheme for JPEG images, where some non-zero AC coefficients valued '1' and '-1' are used as the peak points to embed additional data, and other non-zero AC coefficients shift to make room for additional data while the remaining coefficients keep unchanged. And Huang ${ et~al. }$ took a block selection strategy which is based on the number of zero AC coefficients in the block to decide which block is chosen to embed data first. Good image quality and embedding capacity are realized in \cite{IEEEexample:huang2016reversible}, besides, the file size is kept well. Then Wedaj ${ et~al. }$ \cite{IEEEexample:wedaj2017improved} proposed an improved RDH in JPEG images based on the new coefficient selection strategy, which is improved on the basis of Huang ${ et~al. }$'s work. The additional data is embedded according to the embedding cost in each position of the block. In addition, Hong ${ et~al. }$ \cite{IEEEexample:hong2018improved} and Hou ${ et~al. }$ \cite{IEEEexample:hou2018reversible} also made improvement on the basis of the work of Huang ${ et~al. }$. Hou ${ et~al. }$ \cite{IEEEexample:hou2018reversible} proposed a method based on DCT frequency and block selection. They considered the influence of the quantization step in quantified DCT coefficients' change and simulated distortion in blocks before embedding, the scheme combines the selection strategy of \cite{IEEEexample:huang2016reversible} and \cite{IEEEexample:wedaj2017improved}. The method can keep good visual quality and also keep a small expansion in file size.  Liu ${ et~al. }$ \cite{IEEEexample:liu2018reversible} utilized difference expansion(DE) in RDH scheme for JPEG images which obtains high capacity but the quality of the image is not so good. Because this method modifies the quantized coefficients to a large extent and it causes bigger distortion.

\begin{table}
\caption{The considerations of RDH in JPEG domain.}
\label{tb2considerations of RDH}
\begin{tabular}{|c|c|c|}
\hline
\diagbox{schemes}{considerations} & \multicolumn{1}{|l|}{\begin{tabular}[c]{@{}c@{}}Rate and \\ distortion\\ performance\end{tabular}} & \begin{tabular}[c]{@{}c@{}}File size \\ expansion\end{tabular} \\ \hline
Huang ${et~al. }${\cite{IEEEexample:huang2016reversible}}                        & \checkmark                                                                          &                                                                \\ \hline
Hong ${et~al. }${\cite{IEEEexample:hong2018improved}}                         & \checkmark                                                                           &                                                                \\ \hline
Hou ${et~al. }${\cite{IEEEexample:hou2018reversible}}                          & \checkmark                                                                          &                                                                \\ \hline
Proposed scheme                             & \checkmark                                                                           & \checkmark                                                            \\ \hline
\end{tabular}
\end{table}
\par From all of the above, we can see that image quality and embedding capacity are very important, and file size is equally important for RDH in JPEG images. However, the past all works haven't taken the file size into consideration separately while designing methods. Here, we use file size expansion to better display the changes of the cover file size after embedding data. And Table \ref{tb2considerations of RDH} shows the considerations of RDH in JPEG images while designing the algorithm. It is clear that the state-of-the-art RDHs in JPEG images don't take the file size expansion into account. Thus, we propose a scheme combining not only the rate and distortion performance but also the file size expansion. An optimization strategy is used to balance the two called multi-objective optimization which will achieve different effects according to different requirements. In addition, the distortion and file size expansion of JPEG images are well designed to better map to the image quality and file size. Experiment results show that the method outperforms the previous works in both image quality and file size expansion.

\par In this paper, several sections are displayed in the following. Section \uppercase\expandafter{\romannumeral2} gives the related works which explain some theory acknowledges and some other schemes. And the detail of the proposed scheme is shown in Section \uppercase\expandafter{\romannumeral3}. Section \uppercase\expandafter{\romannumeral4} displays the experimental results, then the paper ends by summarizing our work in Section\uppercase\expandafter{\romannumeral5}.

\section{Related works}\label{Related works}
\par In order to better display our proposed scheme, the prerequisite knowledge is shown first. The compression of the spatial images to the JPEG images can help to well design the RDH scheme by its characteristic or easily comprehend the principle of some RDH schemes in JPEG images. What's more, in our scheme, a multi-objective optimization math model is used to explain the target of RDH in JPEG images: minimize the file size expansion and the distortion simultaneously. Thus, the overview of JPEG compression and the multi-objective optimization are demonstrated in this section. And the based schemes which we improve on are also displayed here.
\subsection{Overview of JPEG Compression}\label{Overview of JPEG compression}
\begin{figure*}
\centering
\includegraphics[width= 7 in]{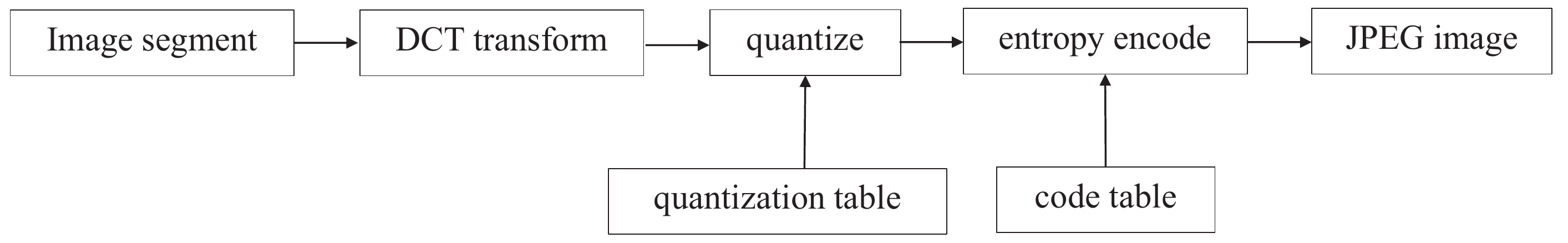}
\caption{The generation process of JPEG image.}
\label{fig1_JPEGgeneration}
\end{figure*}
\par The essence of compression is to remove or reduce redundant information in images, so the file size of compressed images can be small. The procedure of JPEG compression is shown in Fig.~\ref{fig1_JPEGgeneration}. First, the image is divided into nonoverlapping 8${\times}$8 sized blocks. Next, do the DCT transformation on the divided blocks, and this process can be described by (\ref{eq1}).
\begin{equation}
F(u,\!v)\!=\!\frac{1}{4}c(u)c(v)\!\!\!\sum_{x=0}^{7}\!\sum_{y=0}^{7}\!f(x,\!y)\!cos\frac{\!(2x\!+\!1)u\pi}{16}\!cos\frac{\!(2y\!+\!1)v\pi}{16},
\label{eq1}
\end{equation}
where,
\begin{equation}
c(u)=\begin{cases}
 \frac{1}{\!\sqrt{2}},& \text{ if } u= 0\\
 1,& \text{  } otherwise,
\end{cases}
\label{eq2}
\end{equation}
and $(x,y)$ represents the index of a pixel in the spatial image, $f(x,y)$ is the pixel value in the position $(x,y)$, $(u,v)$ is the index of a coefficient in JPEG images, and $F(u,v)$ is the DCT coefficient value in the position $(u,v)$.

\par
\begin{figure}
\centering
\includegraphics[width= 2 in]{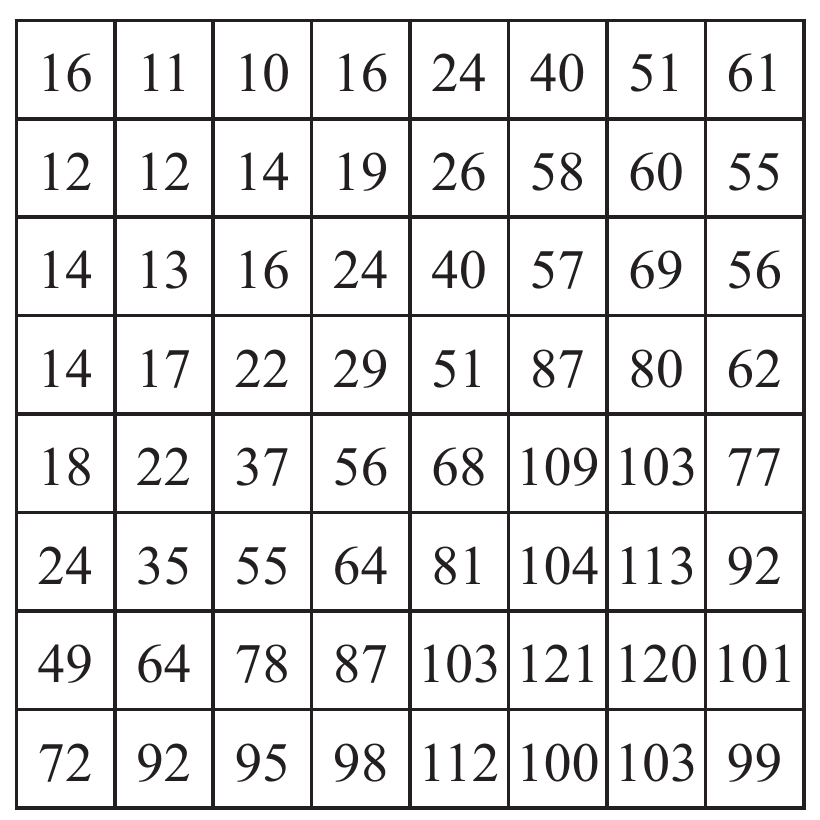}
\caption{The standard quantization table when QF = 50.}
\label{fig2_QF=50}
\end{figure}

 After the DCT transformation, the DCT matrix is gained and the coefficients are called DCT coefficients. Moreover, in each block, the first coefficient is named DC coefficient and the other 63 coefficients are AC coefficients. Then, the quantization is performed on the DCT matrix using the predefined quantization table for different quality factor (QF), the specific operation is that the DCT matrix divides by the quantization table point by point. Fig.~\ref{fig2_QF=50} is a standard quantization table when QF=50, every position in the 8$\times$8 sized block is called a frequency. The process of quantization causes the loss from spatial images to JPEG images, and the QF represents the degree of compression, smaller the QF the more compression. Till now, we get the compressed data, quantized DCT coefficients, which our proposed method makes modification on in this paper. The compression is done, and the quantized DCT coefficients have to be coded. Due to the different characteristics of DC and AC coefficients, different coding methods are taken. Adopt Differential Pulse Code Modulation (DPCM) to encode the DC coefficients because of the correlation of DC coefficients between adjacent blocks, and Run Length Encoding (RLE) to encode AC coefficients, besides, the AC coefficients valued '0' don't need to be encoded. After the DPCM and RLE, the Huffman encoding is further used to acquire the final binary coding of quantized DCT coefficients. And the file size is tightly close to the coding length of quantized DCT coefficients, so keeping the coding length is a good way to keep file size.
\par From the process of compressing the spatial images to the JPEG images, we can clearly see that the DCT procedure changes the distribution of coefficients, and the quantization achieves the essence of compression. Hence, the quantization table helps a lot in reducing the file size expansion while designing the RDH scheme in JPEG images.
\subsection{Multi-objective Optimization}\label{Multi-objective optimization}
\par Multi-objective optimization refers to maximizing or minimizing multiple objectives while designing work, and satisfying constraints. In general, the sub-goals of the multi-objective optimization problem are contradictory. The improvement of one sub-goal may cause the performance of the other sub-goals to be degraded, that is, to simultaneously optimize the multiple sub-goals is not possible, but they can be coordinated and compromised among themselves so that each sub-goal is optimized as much as possible. There are basically the following methods for solving multi-objective optimization: one is to reduce the multi-objective into a single objective which is easier to solve, such as main target method, linear weighting method, ideal point method, etc. The other is called the hierarchical sequence method, that is, the target is given a sequence according to its importance, and each time the next target optimal solution is found in the previous target optimal solution set until the common optimal solution is obtained. It can also be modified by the simplex method. Another method called analytic hierarchy process is a multi-objective decision-making and analysis method combining qualitative and quantitative methods, and it is more practical for the case where the target structure is complex and lacks necessary data.
\par The general multi-objective model is shown as (\ref{eq3}) and (\ref{eq4}), where $X=[x_{1},x_{2},...,x_{n}]^{T}$ is the independent variable, and $f_i (X)$ is the $i$-th objective that needed to be maximized or minimized, $\phi _{j}(X)\leq g_{j}$ is the $j$-th constraint, and there are $n$ variables, $k$ objectives and $m$ constraints in (\ref{eq3}) and (\ref{eq4}). The final goal is to maximize or minimize the $k$ objectives while satisfying $m$ constraints by changing the independent variable $X$. Moreover, the number of variables, objectives and constraints will make great differences in the final result including the running time and the optimized result.
\begin{equation}
F(X)=\begin{pmatrix}
max(min)f_{1}(X)\\
max(min)f_{2}(X)\\
.\\
.\\
.\\
max(min)f_{k}(X)
\end{pmatrix}
\label{eq3}
\end{equation}
\begin{equation}
s.t. \ \Phi (X)=\begin{pmatrix}
\phi _{1}(X)\\
\phi _{2}(X)\\
.\\
.\\
.\\
\phi _{m}(X)
\end{pmatrix}\leq
G=\begin{pmatrix}
g_{1}\\
g_{2}\\
.\\
.\\
.\\
g_{m}
\end{pmatrix}
\label{eq4}
\end{equation}
\subsection{Other Schemes}
\par The multi-objective optimization is employed to minimize the distortion and the file size expansion in RDH for JPEG images, and it is adopted on the basis of other schemes in this paper. Therefore, the other schemes that we improve on are displayed here.
\subsubsection{Huang ${et~al. }$'s scheme}
\begin{figure*}
\centering
\includegraphics[width= 6 in]{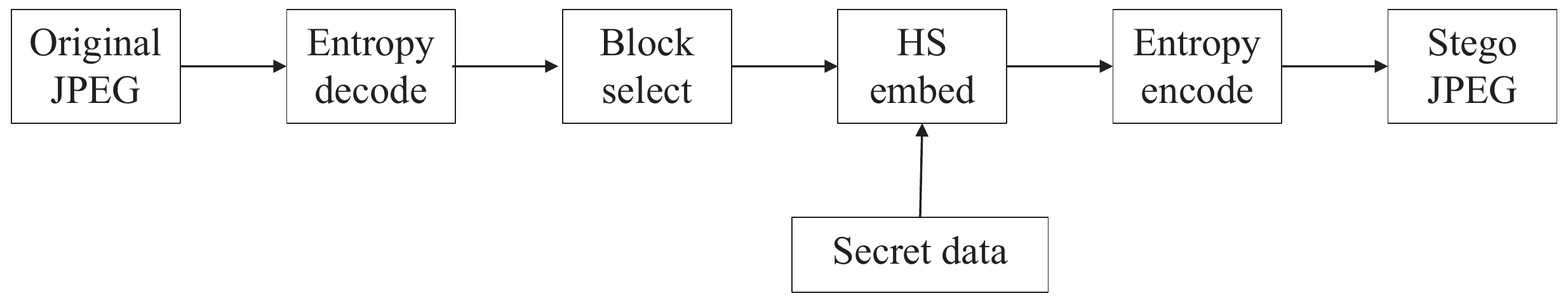}
\caption{The embedding process of Huang ${et~al. }$'s scheme.}
\label{fig3_Huangscheme}
\end{figure*}
\par Huang ${ et~al. }$ proposed a RDH scheme for JPEG images using the existing HS technology, they analyzed the characteristics of DC and AC coefficients and found that the DC coefficients are rather flat in histogram while the AC coefficients are sharper. And another reason they chose the AC coefficients to embed additional data is the changes in DC coefficients may make great distortion in JPEG images because of their containing more information of JPEG images. What's more, it can be seen that in the histogram of AC coefficients, the amplitude of value '0' is the largest and the amplitude of absolute value '1' is the second largest. But for the sake of file size expansion, the AC coefficients valued '0' are not considered to embed data. And then, the AC coefficients valued '1' and '-1' are chosen to embed data combining HS. After the embedded coefficients are determined, which block is selected to embed data first should be taken into account. They selected the block according to the number of AC coefficients valued '0' in the block, the number is smaller, the order of block is later. In general, the block with more AC coefficients valued '0' means flat block, that's to say, there will be less distortion in this block. Following the two stages, coefficients choice and block selection, additional data is embedded into the chosen coefficients in selected blocks. In addition, the embedding process of Huang ${ et~al. }$'s scheme is shown in Fig.~\ref{fig3_Huangscheme}.
\par The scheme achieves good image quality and low file size expansion. And the block selection strategy is proposed first for RDH in JPEG images in \cite{IEEEexample:huang2016reversible}.
Huang ${ et~al. }$'s work made a great push in RDH for JPEG images, but the scheme  only considers the smoothness of blocks, and the number of AC coefficients valued '0' cannot exactly represent the smoothness. From the overview of JPEG compression, we can see that the coefficients in JPEG images are not the same as the one in spatial images, they are got by the process of DCT and quantization. The quantization table has influence on the distortion that distortion will be different for the same coefficients in different positions.
\subsubsection{Hou ${ et~al. }$'s scheme}
\par Hou ${et~al. }$ designed a RDH scheme for JPEG images on the basis of Huang ${ et~al. }$'s work, they made up for the shortcoming of Huang ${ et~al. }$'s work that the quantization table is not taken into consideration. First, this method calculates the average distortion of each frequency in the quantized DCT blocks. Second, select the top $K$ frequencies according to the descending order of the average distortion for embedding. Third, compute the distortion of each block with only $K$ frequencies. Fourth, select the $K$ frequencies of blocks with small distortion to embed data. Finally, do the cycle from second to fourth to choose different $K$ to search for the best PSNR of the stego image.
\par The stage of calculating the average distortion of each frequency in the quantized DCT blocks is the embodiment of considering the quantization step. And the average distortion value is evaluated by (\ref{eq7})-(\ref{eq9}):
\begin{equation}
cost(u,v)=\frac{\sum_{x=0}^{7}\sum_{y=0}^{7}\Delta f(x,y)^{2}}{64},
\label{eq7}
\end{equation}
\begin{equation}
J_{u,v}=(0.5*C_{u,v}+C_{out})*cost(u,v),
\label{eq8}
\end{equation}
\begin{equation}
U\!D_{u,v}=\frac{J_{u,v}}{C_{u,v}},
\label{eq9}
\end{equation}
where $u,v$ represent the position of frequency, $\Delta f(x,y)$ means the modification in the spatial image that caused by the operation in the JPEG domain, and it can be gained by (\ref{eq14}), and $cost(u,v)$ indicates the frequency $(u,v)$'s corresponding distortion in the spatial domain. $C_{u,v}$ is the number of AC coefficients valued '1' and '-1', $C_{out}$ means the number of AC coefficients whose absolute values are bigger than 1, $J_{u,v}$ shows the total distortion of the frequency $(u,v)$ in all blocks, $U\!D_{u,v}$ means the averge distortion of each frequency which is the basis for frequency sorting. What can be seen is that the quantization table is fully utilized in the compute of $cost(u,v)$.
\par It is undeniable that Hou ${ et~al. }$'s method has achieved better effects in image quality and file size expansion. However, the methods of Huang ${ et~al. }$ and Hou ${ et~al. }$ consider only the distortion, and the file size is not considered separately.
Besides, the block selection strategy also can be improved, both of the two methods embed data sequentially in descending order of distortion in blocks. But there may exist some situations that the combination of two blocks with large distortion is better than three combinations with less distortion in the result of distortion and file size. The multi-objective optimization can solve the two disadvantages by its model and the balance of distortion and file size expansion.
\section{Proposed scheme}
\par
\begin{figure*}
\centering
\includegraphics[width= 7 in]{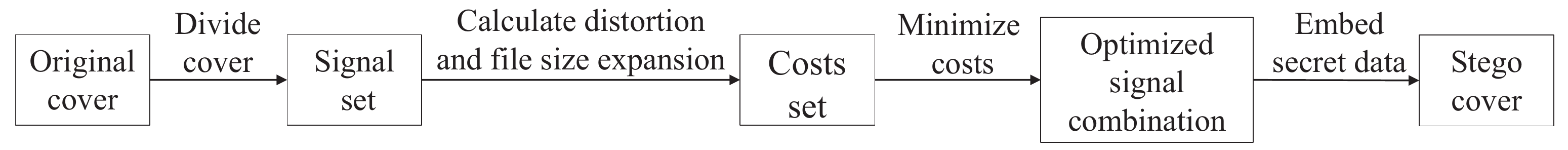}
\caption{The proposed multi-objective optimization embedding process.}
\label{fig4_proposed}
\end{figure*}

 In this section, the proposed scheme is detailed. The framework of the proposed algorithm is shown in Fig.~\ref{fig4_proposed}. We can see from the figure that the how to get the optimized signal combination is the last step before embedding data and it is the core step of our proposed scheme, so the multi-objective optimization is applied in our method to get the optimized signal combination. And the math model suitable for our proposed scheme is described here. Moreover, the costs sets of RDH in JPEG images include three aspects: the embedding capacity, the distortion and the file size expansion. And the embedding capacity is fixed in our proposed scheme for using HS in \cite{IEEEexample:huang2016reversible}, besides, it is significant to better measure the distortion and the file size expansion of the cover. In the following of this section, the calculation of the three costs is described in detail.
\subsection{Math Model}\label{Math model}
\par The metrics of RDH in JPEG images are the payload, image quality, and file size. The ultimate goal of the algorithm is making the image quality best and minimizing the file size while keeping the payload. Thus, the goal in RDH in JPEG images with multi-objective optimization are the image quality and file size, the constraint is the payload. According to the characteristics of objectives in our problems that the file size may decrease as the image quality improve, this paper uses the main target method to solve the multi-objective optimization.
\par In order to facilitate the visual display of the multi-objective optimization formula of RDH in JPEG images, several mathematical parameters are used here to represent the various information of the JPEG image. The cover is divided into $k$ equal-sized blocks to form a signal set $S$, that $S=\begin{Bmatrix}
 s_{1},s_{2},...,s_{k}
\end{Bmatrix}$, here, $s_i$ represents the $i$-th signal. Correspondingly, $R$ is a set of embedding capacity of each signal, $R=\begin{Bmatrix}
 r_{1},r_{2},...,r_{k}
\end{Bmatrix}$, similarly $r_i$ denotes the capacity of the $i$-th signal. $D$ is the image distortion cost set, $D=\begin{Bmatrix}
 d_{1},d_{2},...,d_{k}
\end{Bmatrix}$, and $d_i$ means the distortion of the $i$-th signal after embedding data. $E$ is the file size expansion cost set, $E=\begin{Bmatrix}
 e_{1},e_{2},...,e_{k}
\end{Bmatrix}$, also $e_i$ indicates the $i$-th signal's file size expansion after embedding data.
After JPEG image is represented by these mathematical parameters, the math model can be designed as (\ref{eq5}):
\begin{equation}
\begin{matrix}
\left\{\begin{matrix}
min(V\times E^{T})\\ min(V\times D^{T})
\end{matrix}\right.\\
s.t.\ \ C-V\times R^{T}\leq 0,
\end{matrix}
\label{eq5}
\end{equation}
where $V$ is the decision variable, expressed in the form of $V=\begin{Bmatrix}
 v_{1},v_{2},...,v_{k}
\end{Bmatrix}$, besides, $ v_{i}\in \left \{ 0,1 \right \} $, and if $v_i$ equals 0, the $i$-th signal will not be used to embed data, in the contrast, if $v_i$ equals 1, the $i$-th signal will be selected for embedding. In addition, the $V \times E^T$ represents the total file size expansion, the sum of the file size expansion in the selected signals. And $V \times D^T$ means the total distortion, the sum of distortion after embedding in the selected signals. In addition, $min$ is the minimization function making our goal minimized, $C$ is the length of the additional data and $V \times R^T$ represents the total embedding capacity of the selected signals, and $C-V \times R^T  \leq 0$ is the constraint that the selected signals must satisfy the payload.
\begin{equation}
\begin{matrix}
min\ \  V\times D^{T}\\
s.t.\left\{\begin{matrix}
C-V\times R^{T}\leq 0\\
V\times E^{T}\leq E^{*}+\alpha E^{*},
\end{matrix}\right.
\end{matrix}
\label{eq6}
\end{equation}
\par To better solve this multi-objective problem and reduce the computational complexity, we convert one of the goals into a constraint like (\ref{eq6}). The model in (\ref{eq6}) means that we convert the file size expansion into a constraint and minimize the distortion, where $E^*$  is the optimal value of the target file size expansion, and the $E^*+\alpha E^*$ is the limit of file size expansion in the model and the weight $\alpha$ means the degree of constraint that the file size expansion needs to satisfy. Applying (\ref{eq6}), the multi-objective optimization problem in the proposed scheme can be solved very well.
\subsection{Costs Getting}
\subsubsection{embedding capacity}
\par The embedding capacity of each signal is the number of data each signal can embed, and it is depending on the embedding method. If the adopted embedding method is HS in AC coefficients, then the embedding capacity is the number of AC coefficients which are selected to embed, usually valued '1' and '-1'.
All the coefficients that can embed data are embeddable coefficients. In our proposed scheme, HS is utilized to embed data as Huang $et~al.$'s scheme, so the signal embedding capacity is the number of AC coefficients valued '1' and '-1' in each signal.
\begin{equation}
r_i=\sum_{j=2}^{64} sign(s_{ij}),
\label{eq10}
\end{equation}
\begin{equation}
sign(s_{ij})=\begin{cases}
1,&\text{ if } s_{ij} \ is \ embeddable \ coe\!f\!ficient \\
0,&\text{ if } s_{ij} \ isn't \ embeddable \ coe\!f\!ficient.
\end{cases}
\label{eq11}
\end{equation}
\par The embedding capacity $r_i$ of each signal $s_{i}$ is computed by (\ref{eq10}), where $s_{ij}$ represents the $j$-th coefficient in the $i$-th signal, and $sign(s_{ij})$ means whether $s_{ij}$ is embeddable or not.
\subsubsection{distortion designing}

\par
\begin{figure*}
\centering
\includegraphics[width= 6 in]{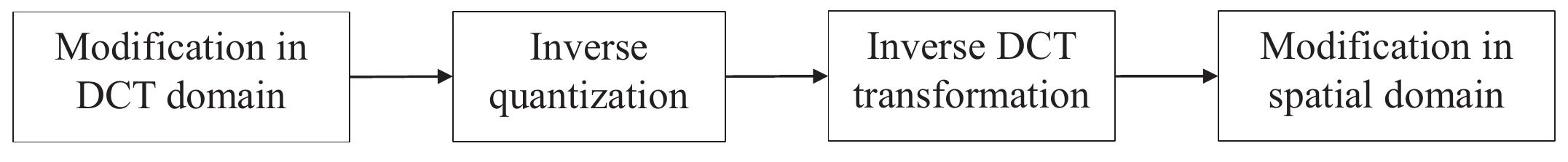}
\caption{The distortion calculation in the spatial domain caused by changes in the DCT domain.}
\label{fig5_distortion}
\end{figure*}

 Image quality is a criterion to measure the performance of the RDH in JPEG images, so the distortion function is designed to reflect the image quality which is inversely proportional to it. And the image quality is reflected in the spatial domain, but we modify the coefficients in the DCT domain to embed data. Therefore, if we want to detect the distortion in spatial images using the modification in quantized DCT coefficients, we must establish a link from the DCT domain to the spatial domain for distortion. We can analyze from the generation process of JPEG images shown in section \ref{Overview of JPEG compression} that the inverse quantization and the inverse DCT transformation should be executed in order on the modification of the quantized DCT coefficients to exactly reflect the modification in spatial images. And the inverse process is displayed in Fig.~\ref{fig5_distortion}. According to the theoretical analysis, the distortion function in \cite{IEEEexample:hou2018reversible} is designed very well, so we use it as our distortion function to measure the image quality, and it is shown following:
\begin{equation}
d_i=I\!DCT(di\!f\!f.*Q),
\label{eq12}
\end{equation}
\begin{equation}
di\!f\!f=s_i-s_i',
\label{eq13}
\end{equation}

\par Assume that $s_i'$ is the current signal that emulates embedding the additional data, $s_i$ is the original signal, $di\!f\!f$ represents the modification of quantized DCT coefficients due to the embedding of additional data. The operator $.*$ means multiplying point by point. $Q$ displays the quantization step, and $di\!f\!f.*Q$ is the process of inverse quantization. IDCT is the function that does inverse DCT transformation, the compute of IDCT transformation is as (\ref{eq14}):
\begin{equation}
f(x,\!y)\!=\!\frac{1}{4}\!\sum_{u=0}^{7}\!\sum_{v=0}^{7}\!\!c(u)c(v)\!F(u,\!v)\!cos\frac{\!(2x\!+\!1)u\pi}{16}\!cos\frac{\!(2y\!+\!1)v\pi}{16},
\label{eq14}
\end{equation}
where $F(u,v)$ represents the coefficient in quantized DCT matrix, $(u,v)$ is the index of position of the coefficient in quantized DCT matrix, and $c(u)$, $c(v)$ is the same meaning as that in (\ref{eq2}). Besides, $(x,y)$ is the index of position of the pixel in the spatial domain. After two inverse transformations, the changes in the DCT domain are well mapped to the spatial domain, and then it can well reflect the quality of stego images.
\subsubsection{file size designing}
\par The significance of file size expansion during embedding is obvious from the analysis in section \ref{Introduction} and \ref{Math model}. But the file size has never been considered separately in the design of existing RDH in JPEG images. How to design the file size expansion of a signal is very important.
\begin{equation}
e_i=\frac{L(s_i')-L(s_i)}{L(s_i)}\times 100\%,
\label{eq15}
\end{equation}
\par The file size expansion can be calculated by (\ref{eq15}), where $s_i$ is the $i$-th signal in the cover, and $L(s_i)$ is the function to count the coding length of the signal $s_i$. And $e_i$
means the ratio of the stego signal file size to the original signal file size in each signal $s_i$, which is the percentage increase of the file size after embedding. We use the percentage increase of the file size instead of the file size to measure the effects of the scheme, that is because the percentage can intuitively reflect the file size changes in the original signal.

\subsection{Data Embedding and Extraction}
\par There will give the processes of data embedding and extraction, and the embedding method is the HS which is the same as the method in \cite{IEEEexample:huang2016reversible}, where the AC coefficients valued '1' and '-1' are used to embed additional data, if the data is '0', the embeddable coefficients keep unchanged, if the data is '1', the embeddable coefficients move to a direction with a large absolute value by 1. The other AC coefficients whose absolute values are bigger than the embeddable coefficients shift to a direction with a large absolute value by 1, and the AC coefficients valued '0' is immovable.
\subsubsection{Data embedding}\label{Data embedding}
\begin{itemize}
\item Get the quantized DCT coefficients from the original JPEG image by decoding.
\item Divide the DCT coefficients into some non-overlapping signals of the same size according to different schemes. The size of the signal in this paper is 8${\times}$8, but it can be any size.
\item Compute the corresponding distortion cost, file size expansion and embedding capacity of each signal in the signal set applying the (\ref{eq10})-(\ref{eq15}).
\item Adopt the multi-objective optimization proposed in section \ref{Math model} to generate the decision variable matrix $V$ used to guide which signal is selected to embed data into. Notice that the multi-objective optimization function must satisfy the given payload, meanwhile, the distortion and file size expansion are optimal.
\item Embed the additional data into the signals according to the decision variable matrix $V$.
\item Encode the modified JPEG image to get the stego image.
\end{itemize}
\par What has to be aware of is that the decision variable matrix $V$ is needed to be compressed, and then it replaces the LSBs of some special signals in the signal set like the first or the last signal, next, the LSBs of these signals are embedded following the additional data.
\subsubsection{Data extraction}
\begin{itemize}
\item First, the quantized DCT coefficients are gained from the stego JPEG image in the same way as the first step in section \ref{Data embedding}.
\item Perform the same block processing on the quantized DCT coefficients as embedding to obtain signals of the same size that do not overlap.
\item Extract the data and recover the signals according to the decision variable matrix $V$. And the decision variable matrix $V$ is extracted first from the LSBs used in the embedding process.
\item Encode the recovered signals again to get the recovered JPEG image which is the same as the original JPEG image.
\end{itemize}
\section{Experimental results}\label{Experimental results}
\par
\begin{figure*}
\centering
    \subfigure[]{
    \includegraphics[width= 1.7 in]{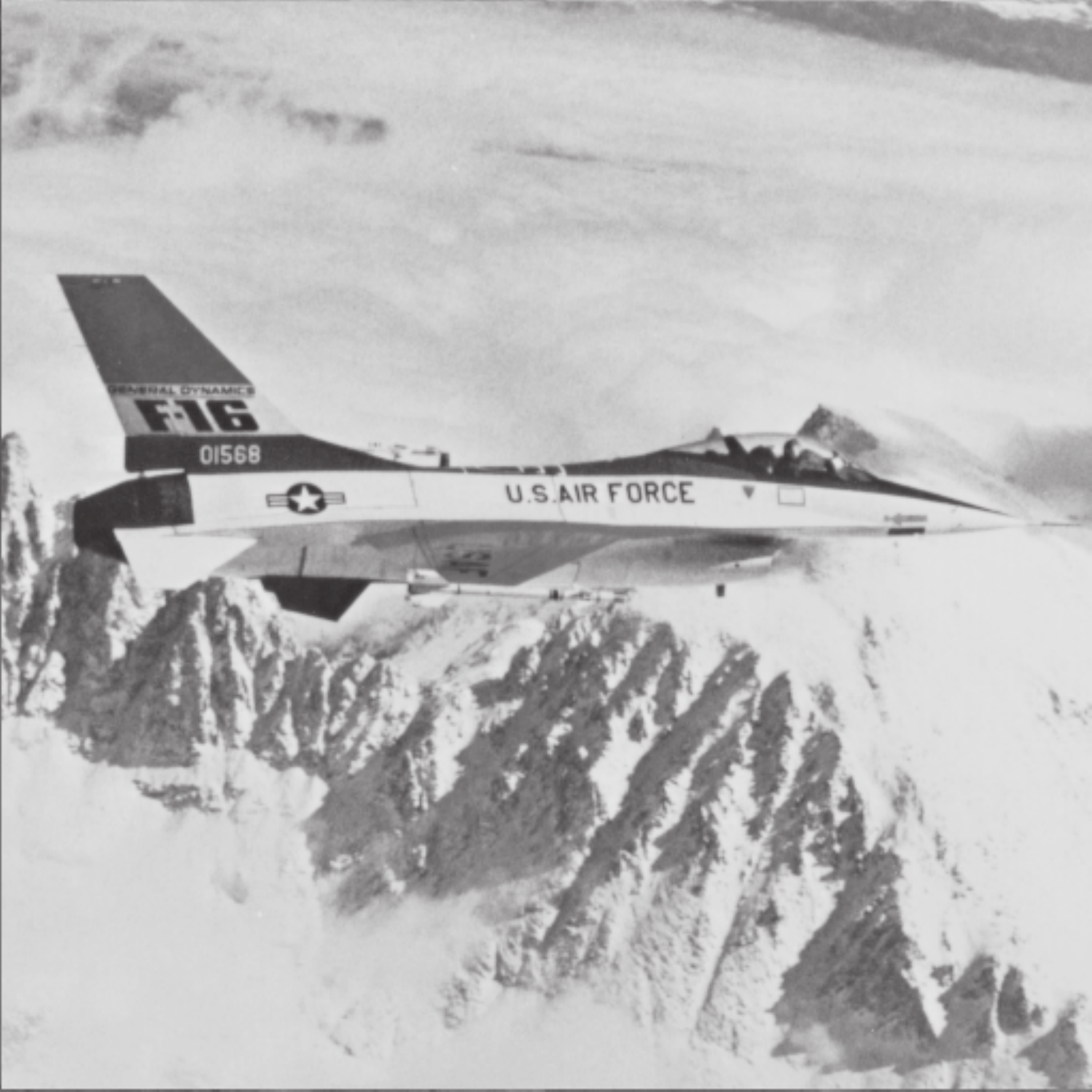}}
    \subfigure[]{
    \includegraphics[width= 1.7 in]{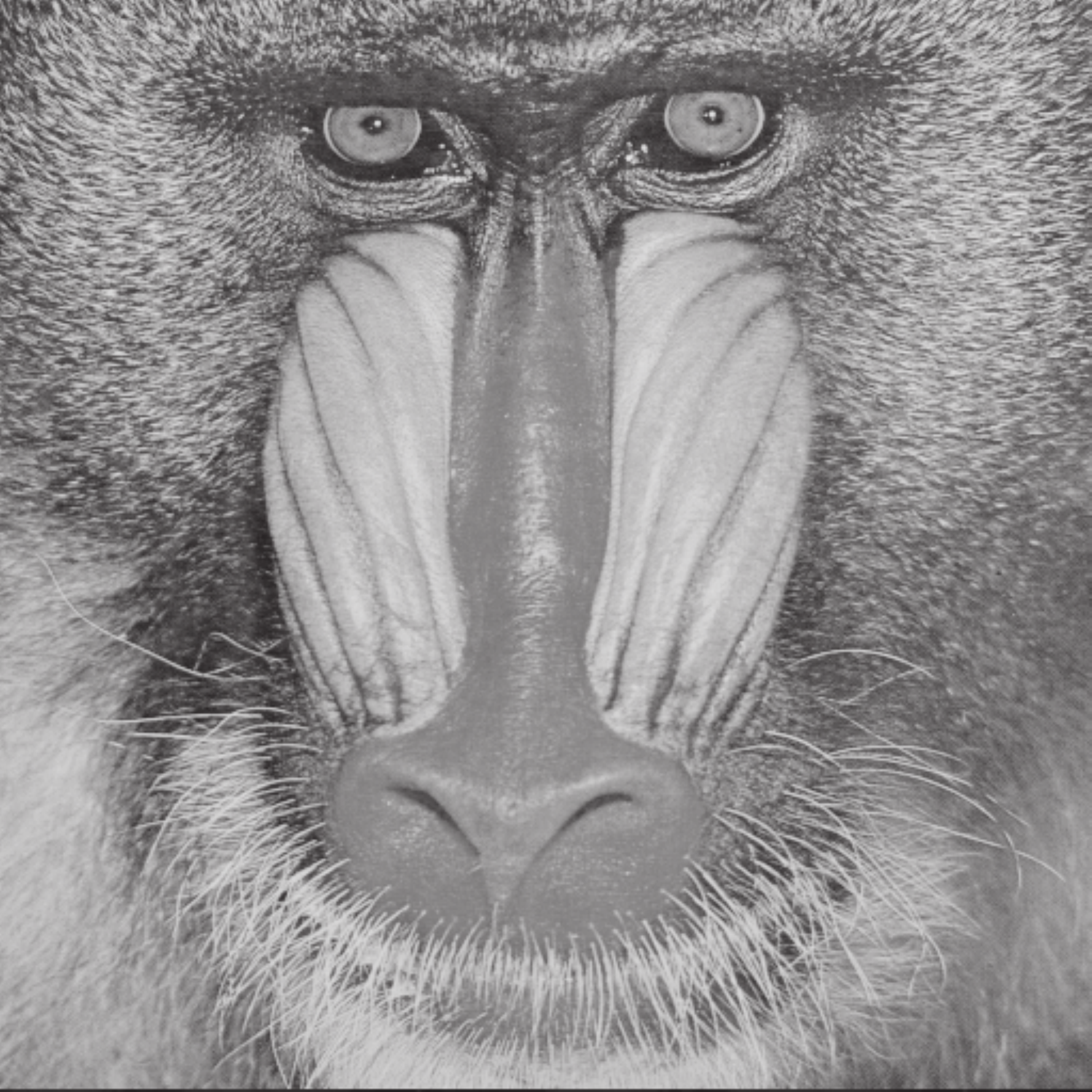}}
    \subfigure[]{
    \includegraphics[width= 1.7 in]{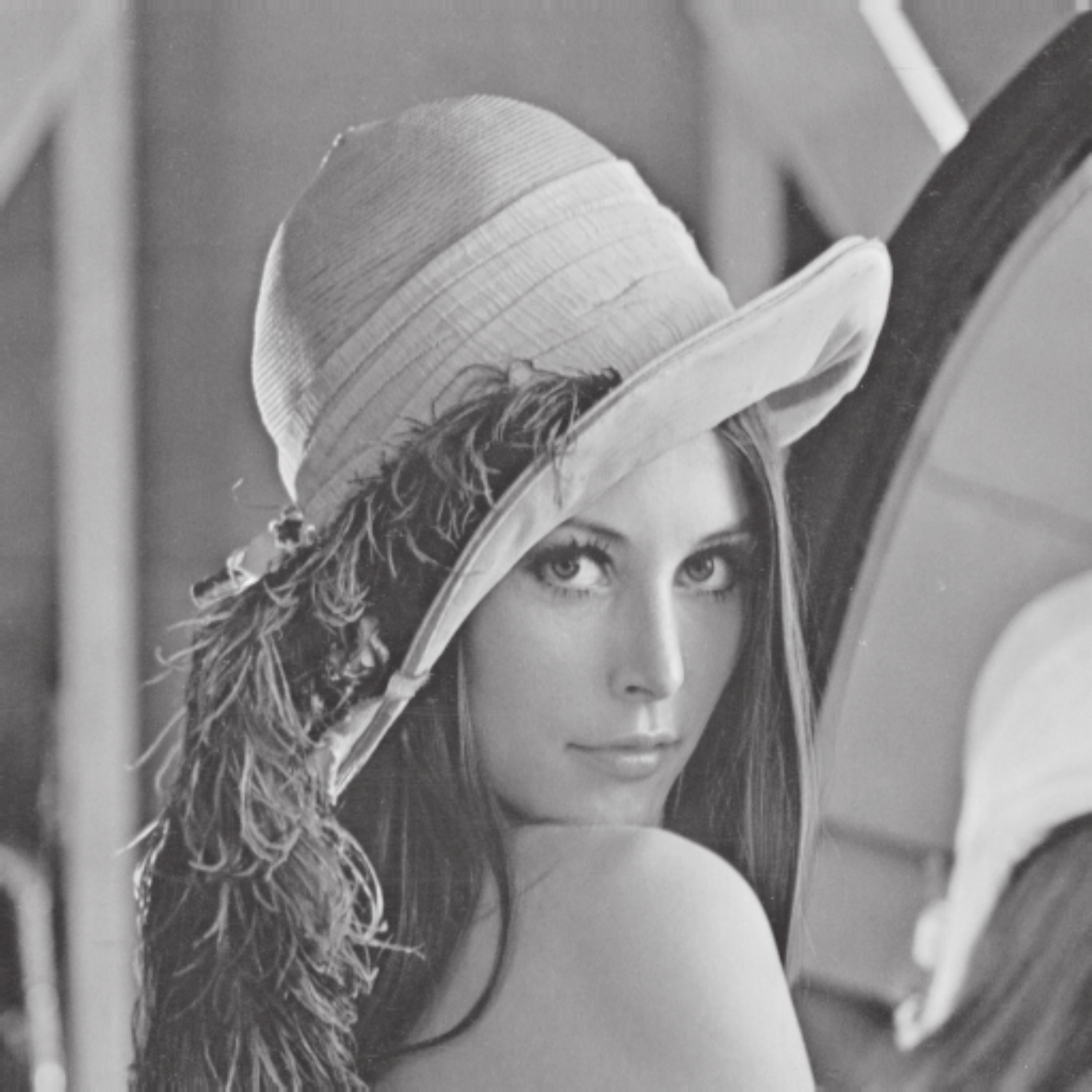}}
    \subfigure[]{
    \includegraphics[width= 1.7 in]{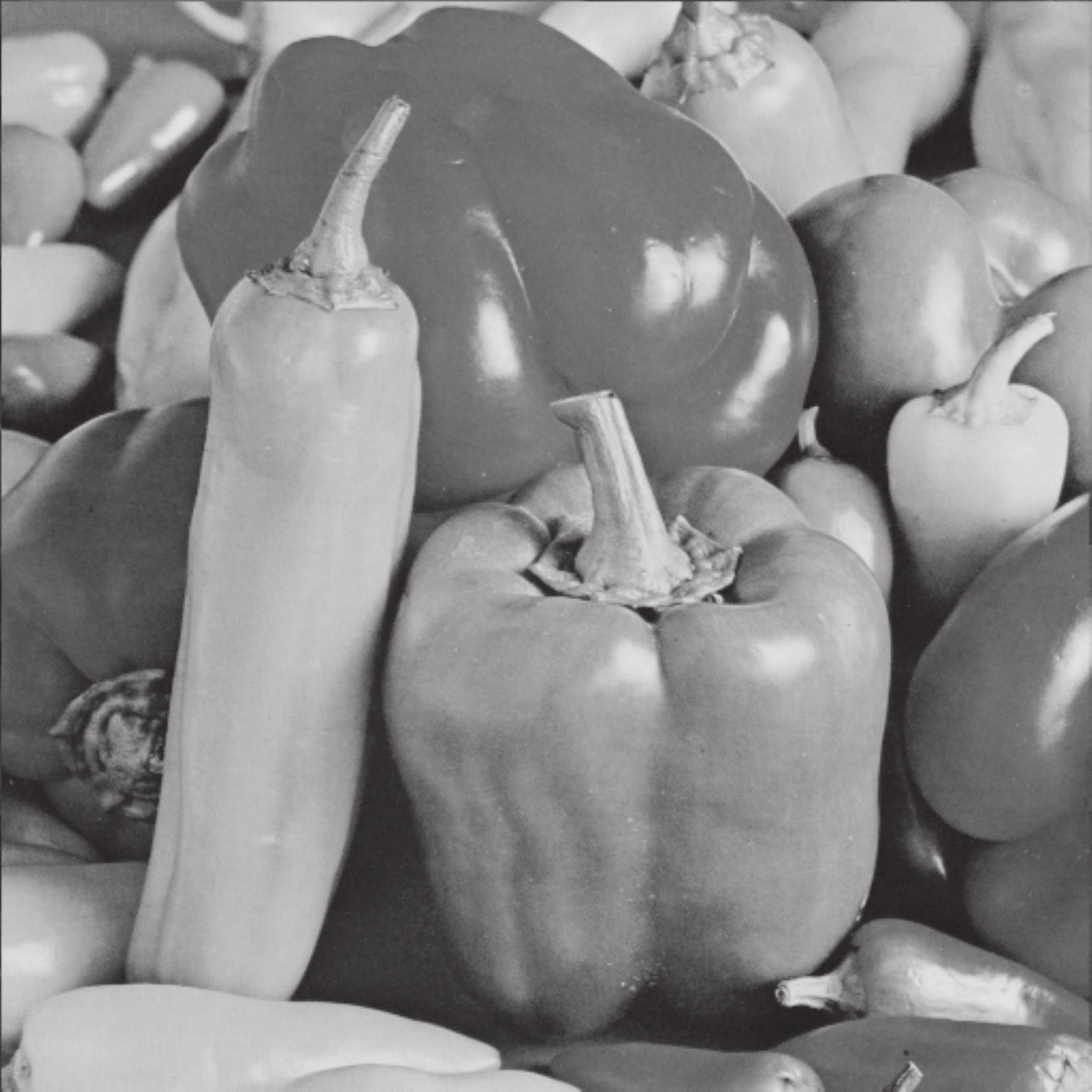}}
\caption{Four test images sized 512${\times}$512. (a)Airplane. (b)Baboon. (c)Lena. (d)Peppers.}
\label{fig6_testimage}
\end{figure*}

 In this section, we display the better performance of our proposed scheme in the image quality and the file size expansion based on Huang ${ et~al. }$'s scheme and Hou ${ et~al. }$'s scheme. Four gray images with size of 512${\times}$512, Airplane, Baboon, Lena, Peppers shown in Fig.~\ref{fig6_testimage} are used to produce the cover signal set in our experiment. And they are compressed to JPEG images with different quality factors (QF=30,50,70,90), and also 96 gray images \cite{IEEEexample:g512} sized 512${\times}$512 are applied to test the average performance of experiments. Besides, the weight $\alpha$ in (\ref{eq6}) we used is set to 1, if it is too small, the optimal decision variable matrix cannot be gained, because the file size expansion is limited to a too small range, and if the weight is large, the optimal decision variable matrix will keep unchanged. Therefore, we choose 1 as the optimal weight according to the best result of our experiments on different weights. We discuss the performance of the algorithm as following.
\begin{figure*}
\centering
    \includegraphics[width= 3.2 in]{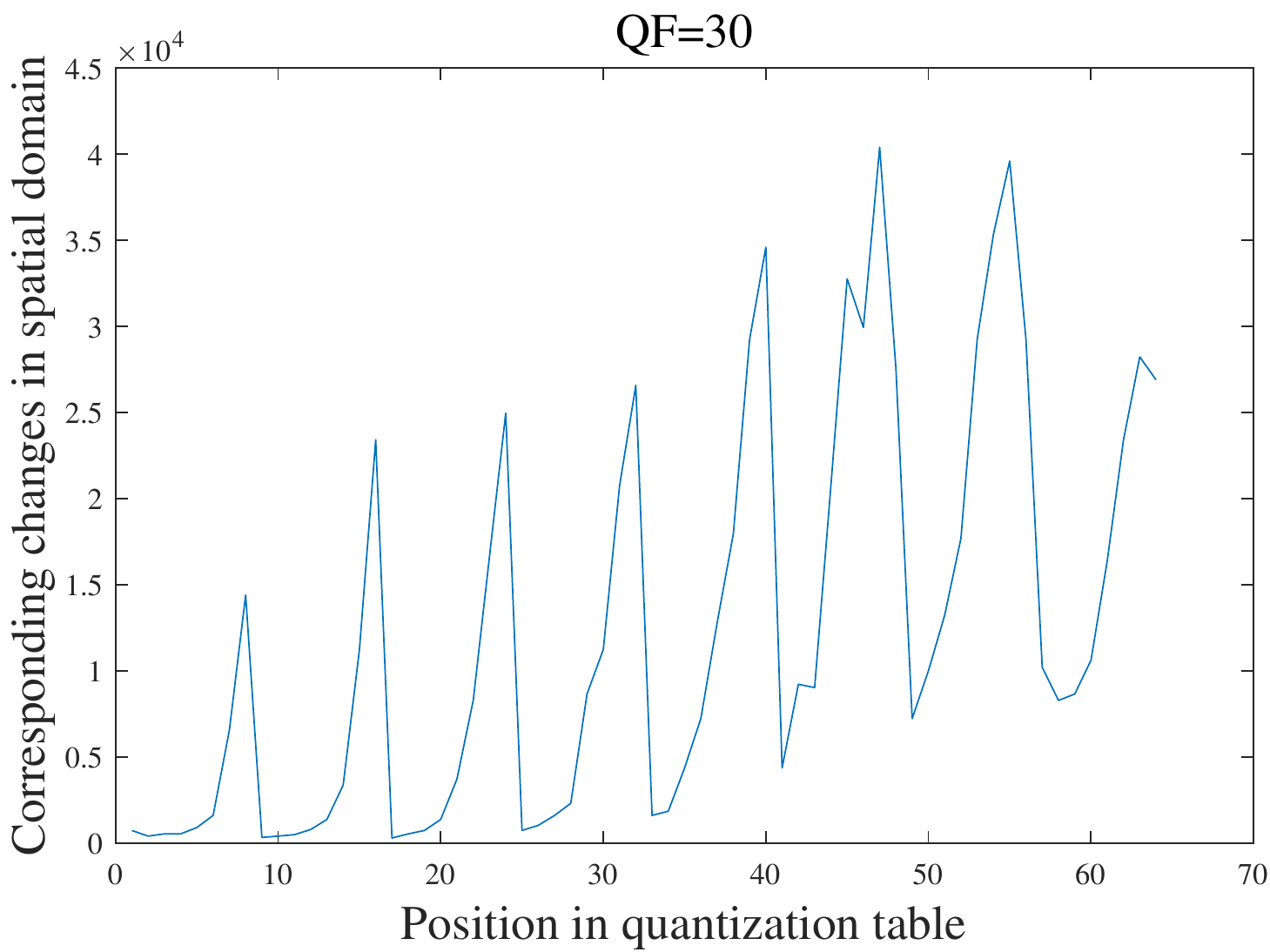}
    \ \ \
    \includegraphics[width= 3.2 in]{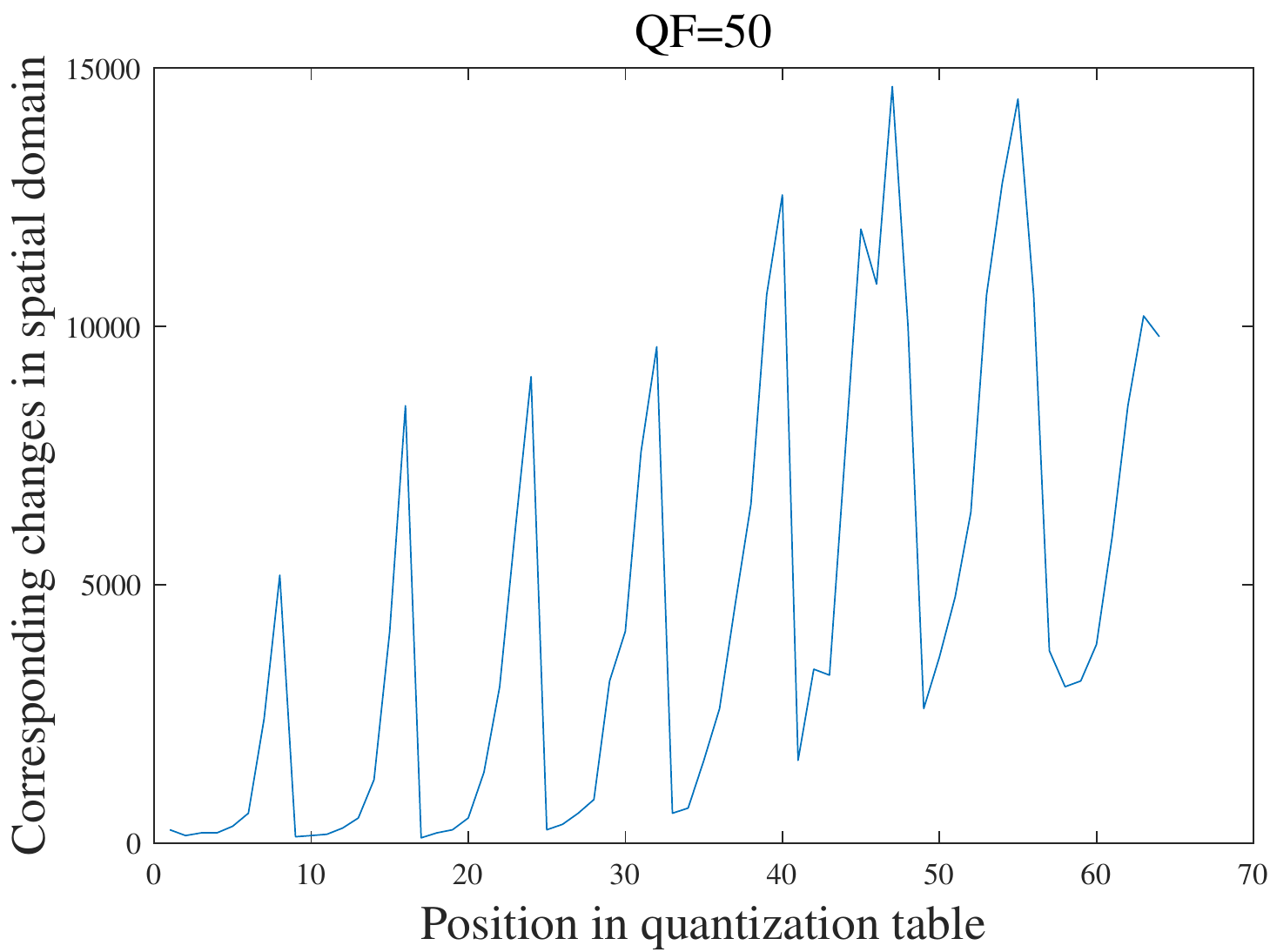}
    \includegraphics[width= 3.2 in]{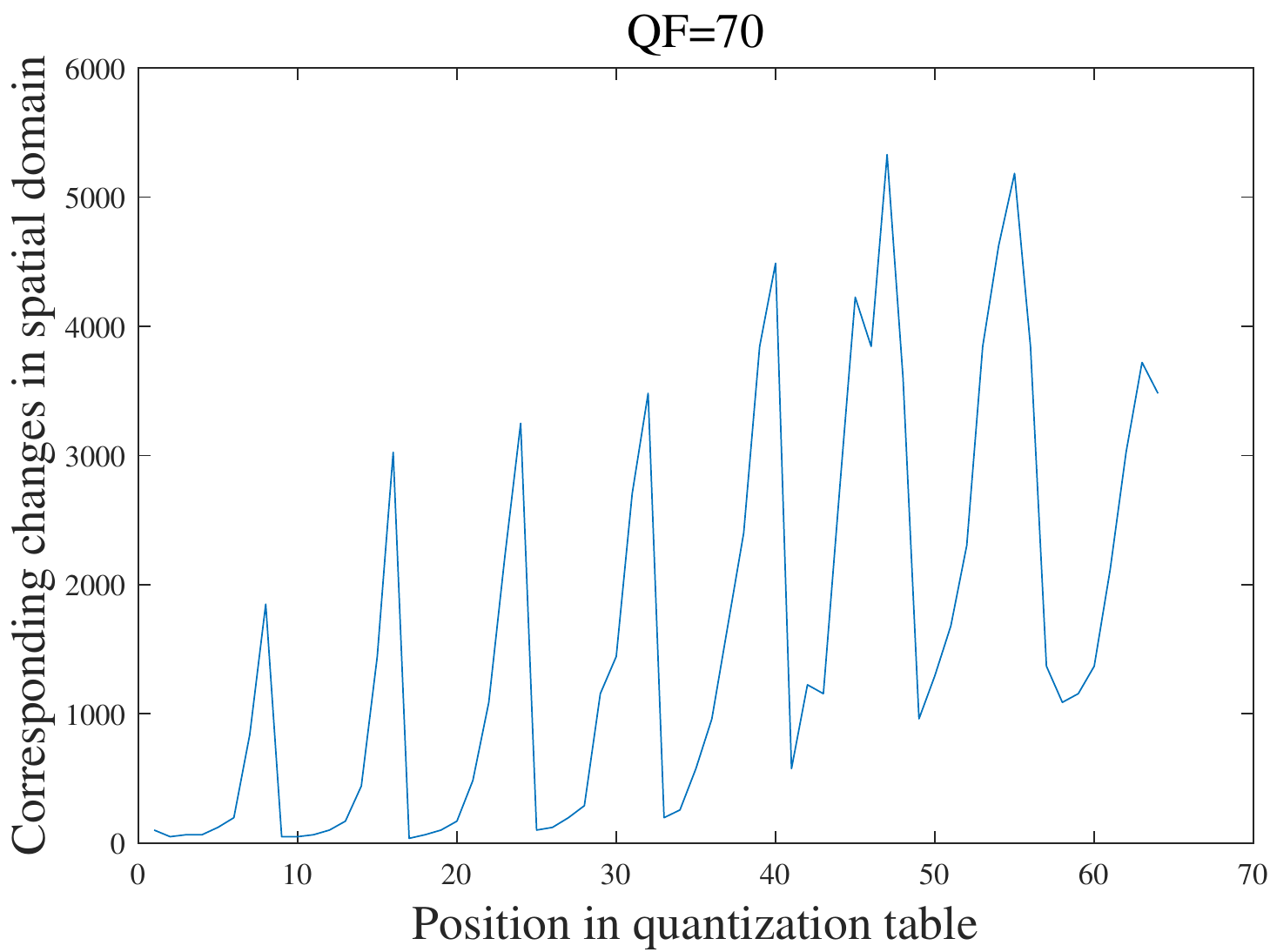}
    \ \ \
    \includegraphics[width= 3.2 in]{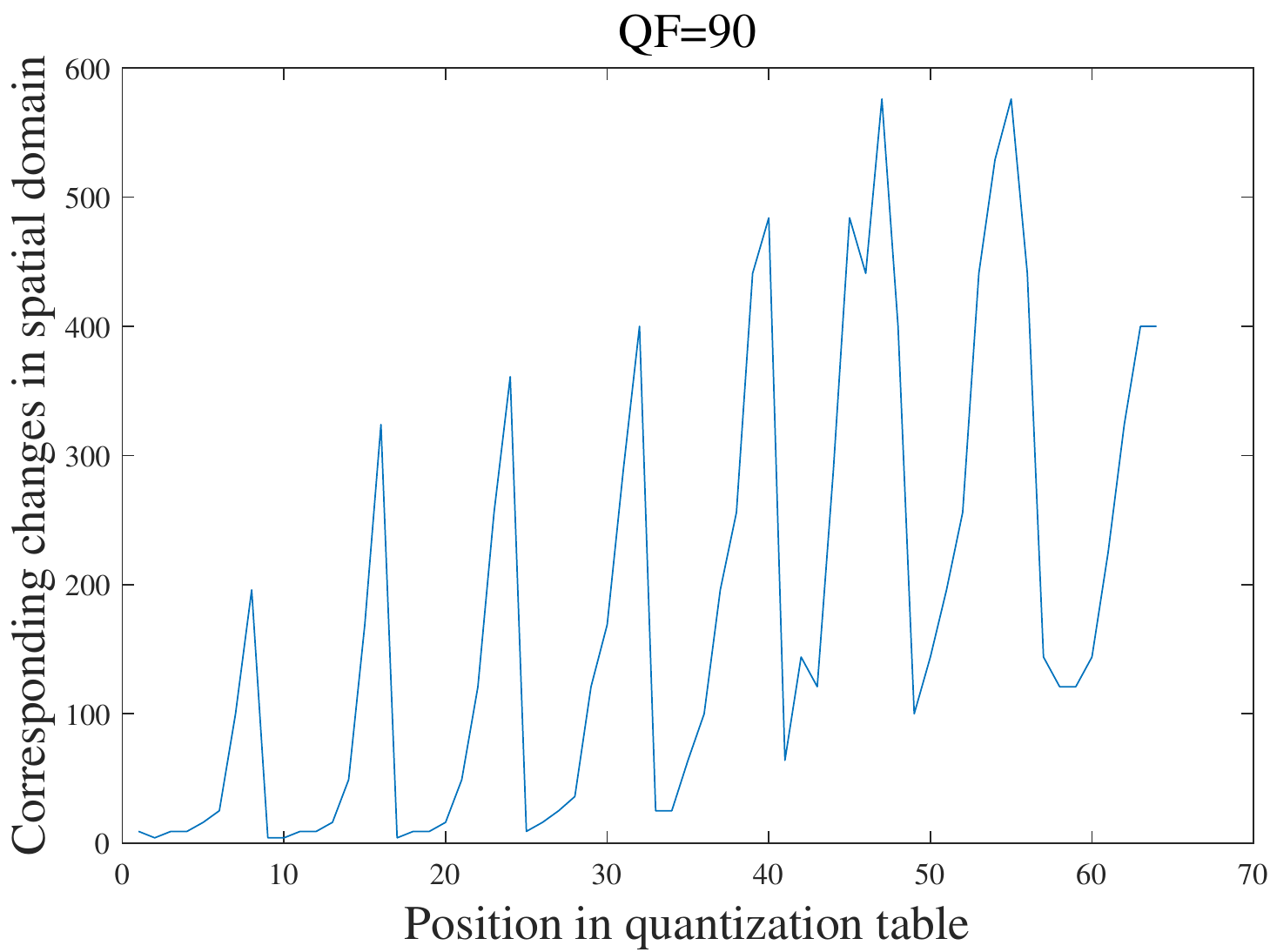}
\caption{The changes of coefficients in spatial domain while modifying one-bit in different frequencies of the quantization table.}
\label{fig7_QFinfluence}
\end{figure*}
\subsection{Influence of quantization step}
\par The impact of quantization step has been mentioned in the previous section. Fig.~\ref{fig7_QFinfluence} gives the influence of quantization step in different quantization tables, different QFs correspond to different standard quantization tables. The abscissa indicates the position of the quantization step in quantization table, and it changes from 1 to 64, which means there are 64 quantization steps in the 8${\times}$8 DCT matrix. The ordinate represents the corresponding change in the spatial domain if there is one-bit change of DCT coefficient in one position. We test the influence of quantization step using the modification in the spatial domain which can be mapped by the corresponding change in the DCT domain. The DCT coefficient of only one position is changed by one bit at a time, and then the corresponding change in the spatial domain is calculated by the inverse quantization and IDCT transformation which is shown in (\ref{eq12})-(\ref{eq14}). The influences of the quantization step are clearly to be seen from Fig.~\ref{fig7_QFinfluence}, the low amplitude means fewer changes in the spatial domain, and it indicates the less distortion. The different influences of different positions are evident, so the quantization step is applied in our distortion function designing to well reflect the changes in spatial domain which are caused by embedding additional data in the DCT coefficients. From the general trend, we can observe that the lower the position, the smaller the distortion will be.
\begin{figure*}
\centering{
    \includegraphics[width= 3.1 in]{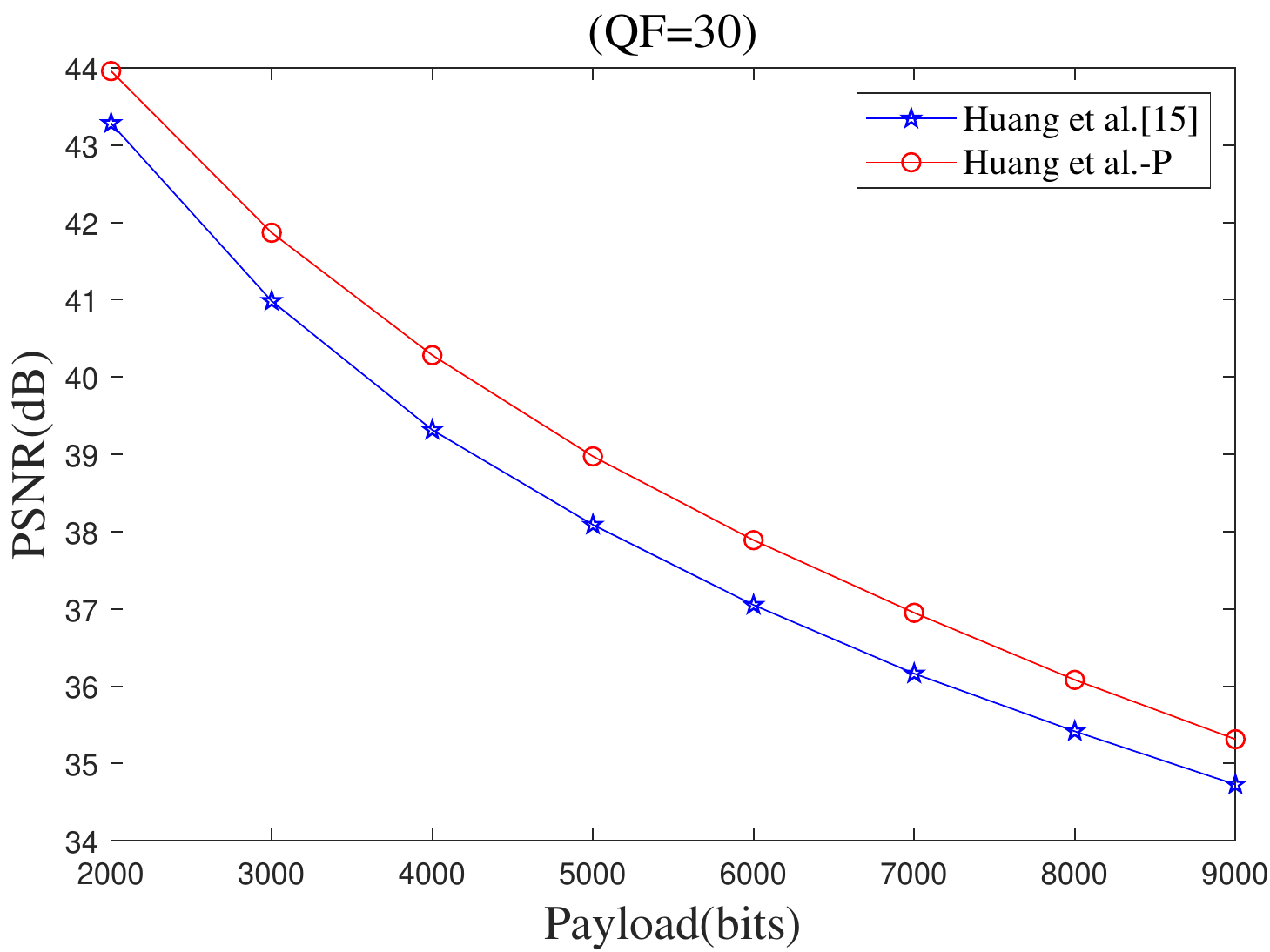}
    \includegraphics[width= 3.1 in]{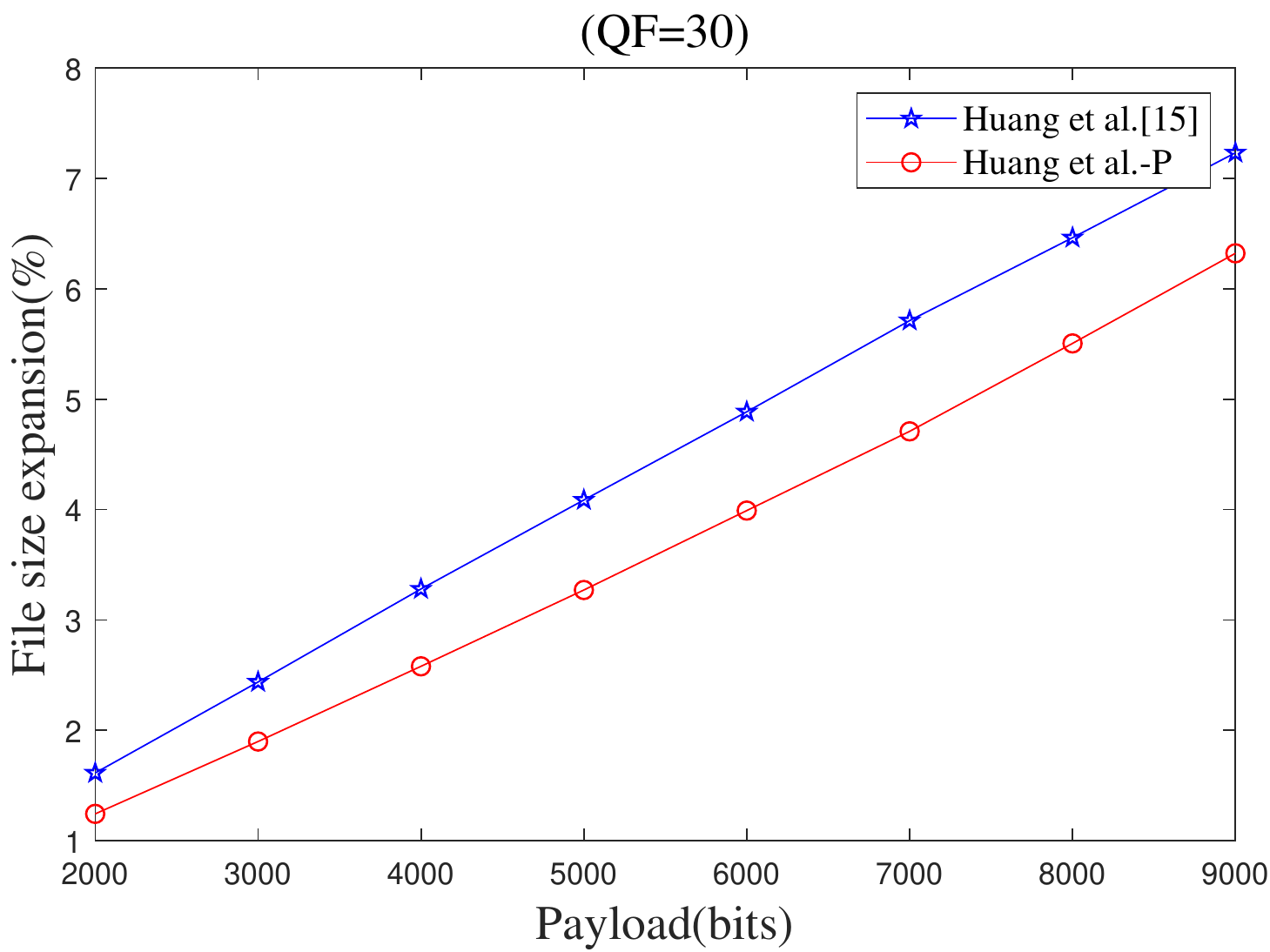}
    \includegraphics[width= 3.1 in]{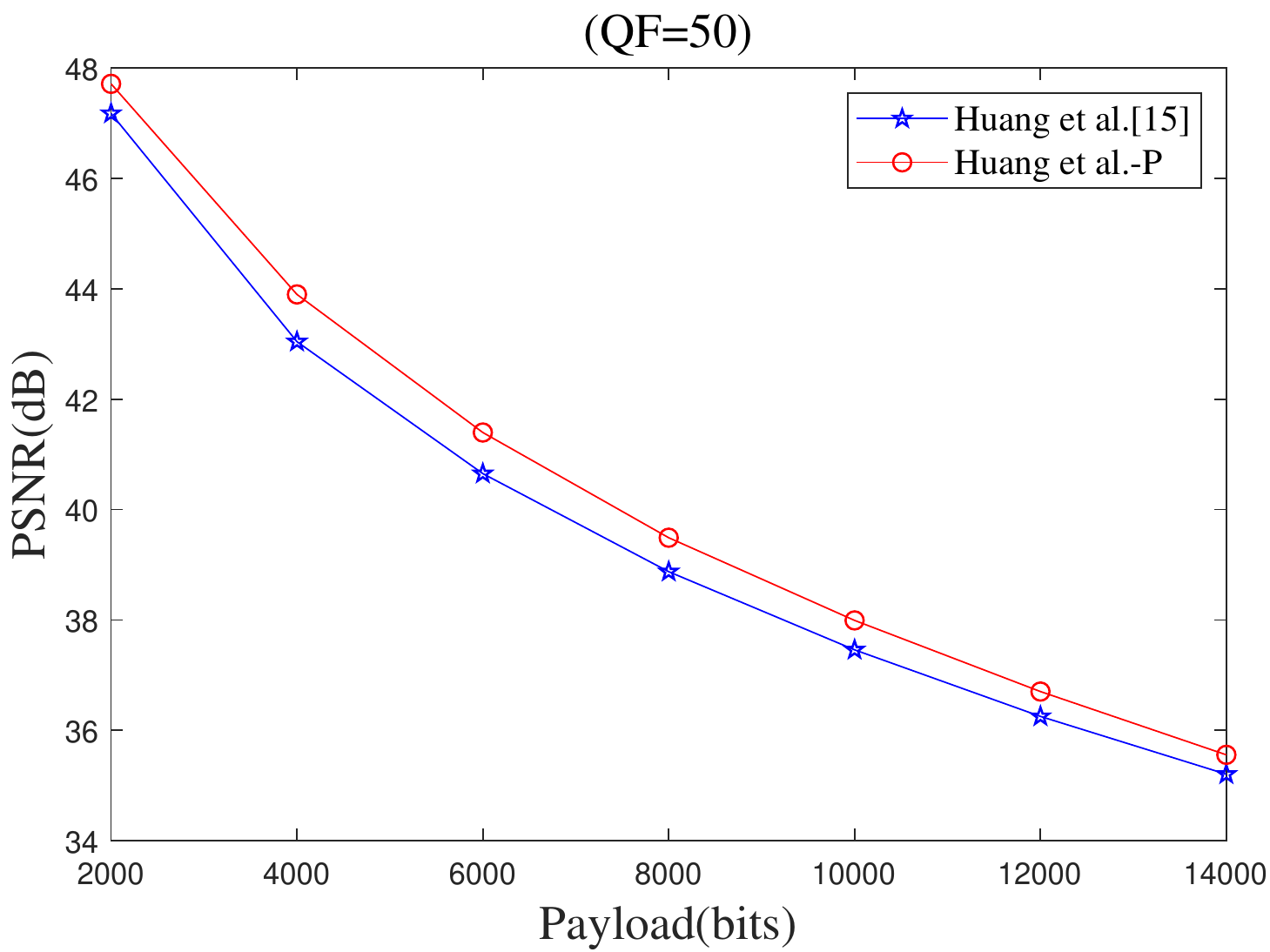}
    \includegraphics[width= 3.1 in]{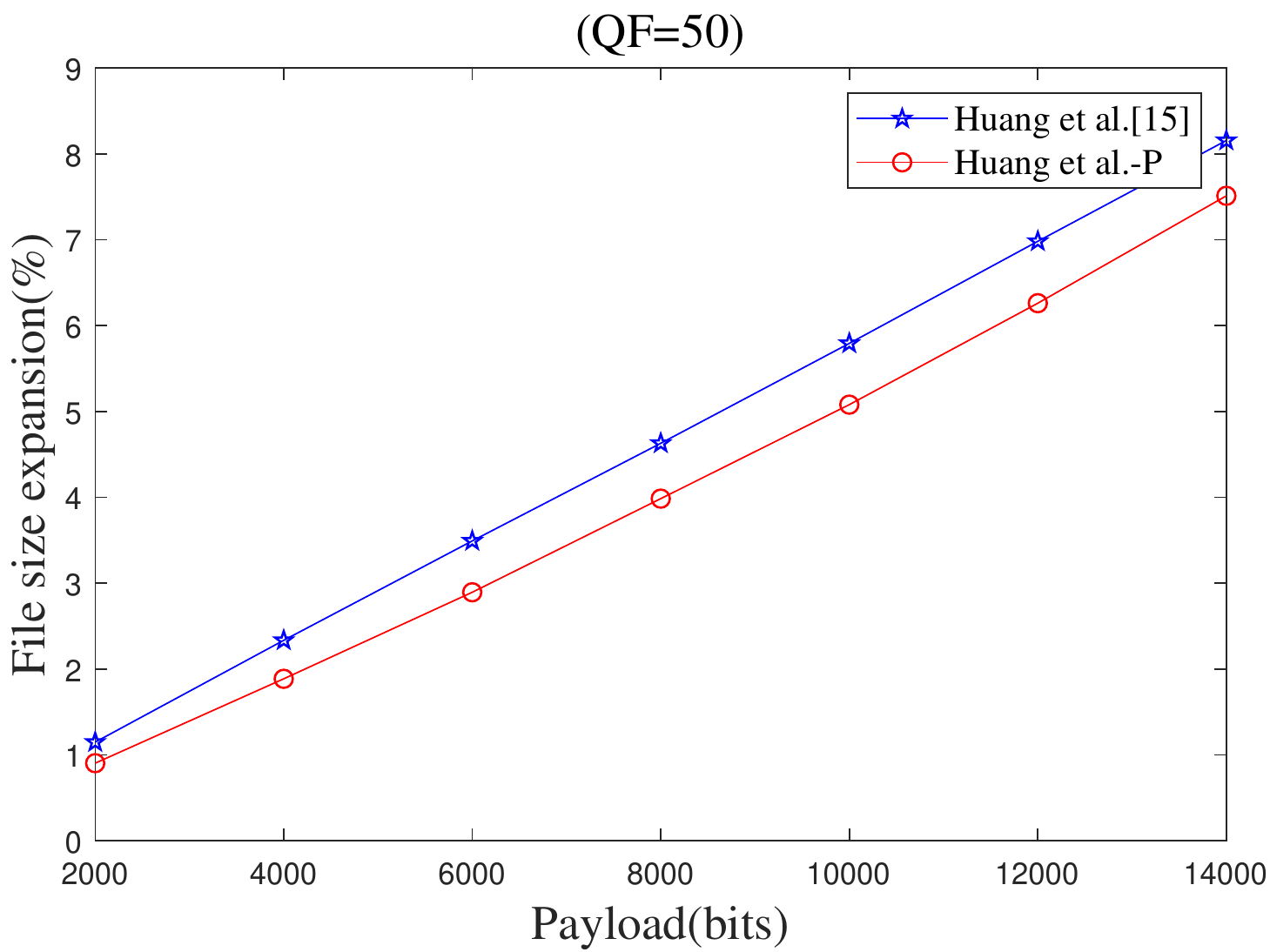}
    \includegraphics[width= 3.1 in]{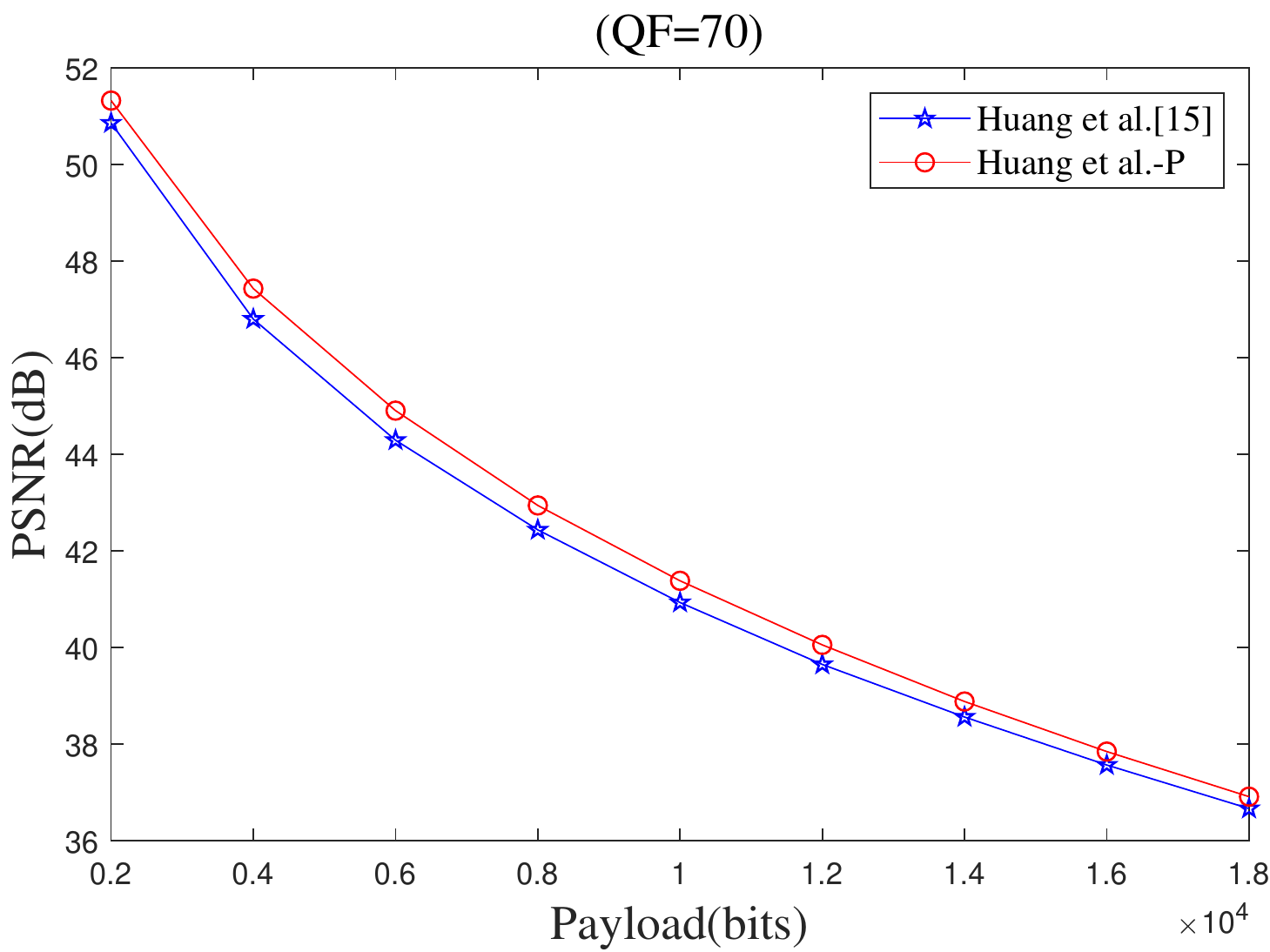}
    \includegraphics[width= 3.1 in]{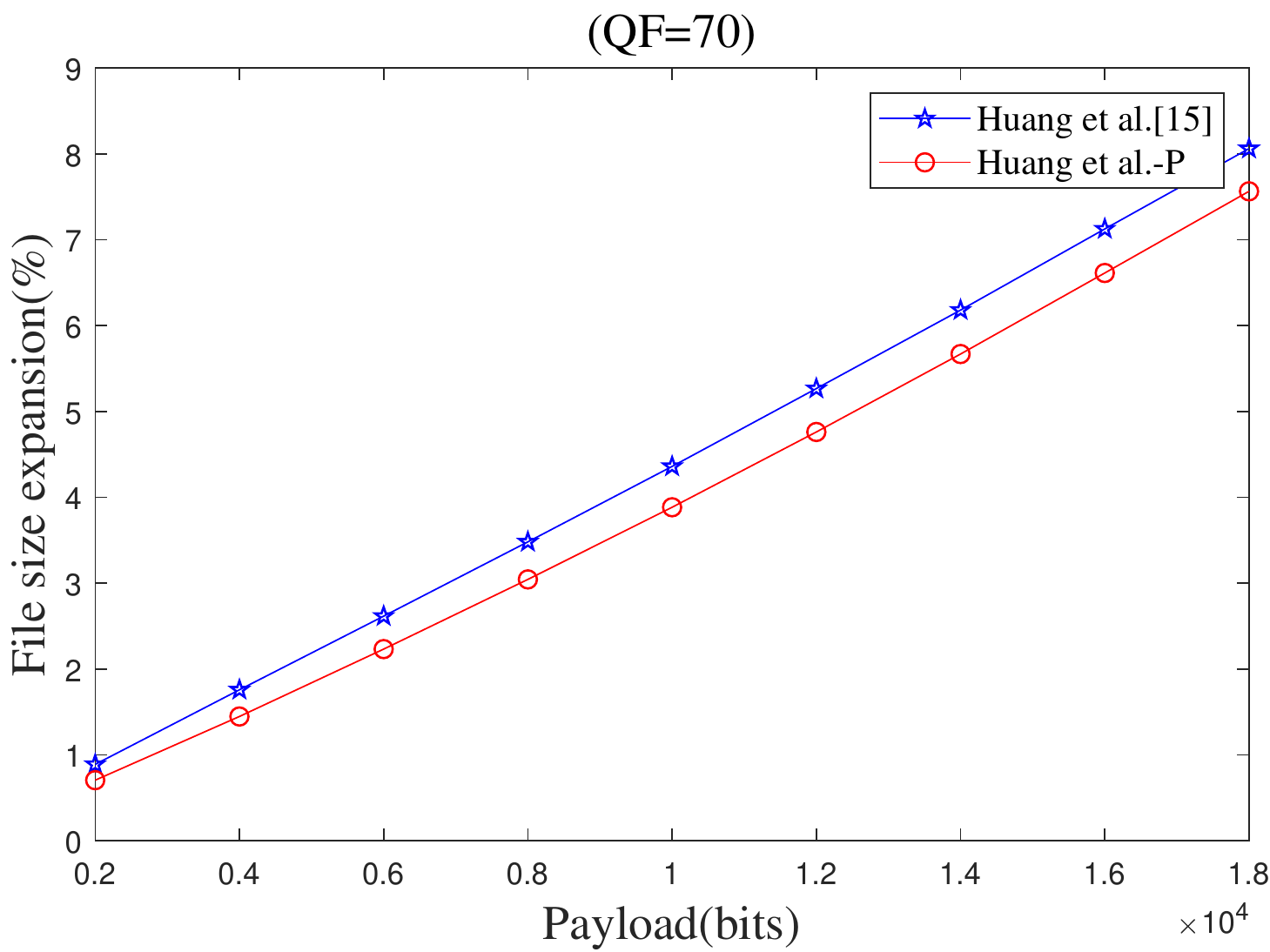}
    \includegraphics[width= 3.1 in]{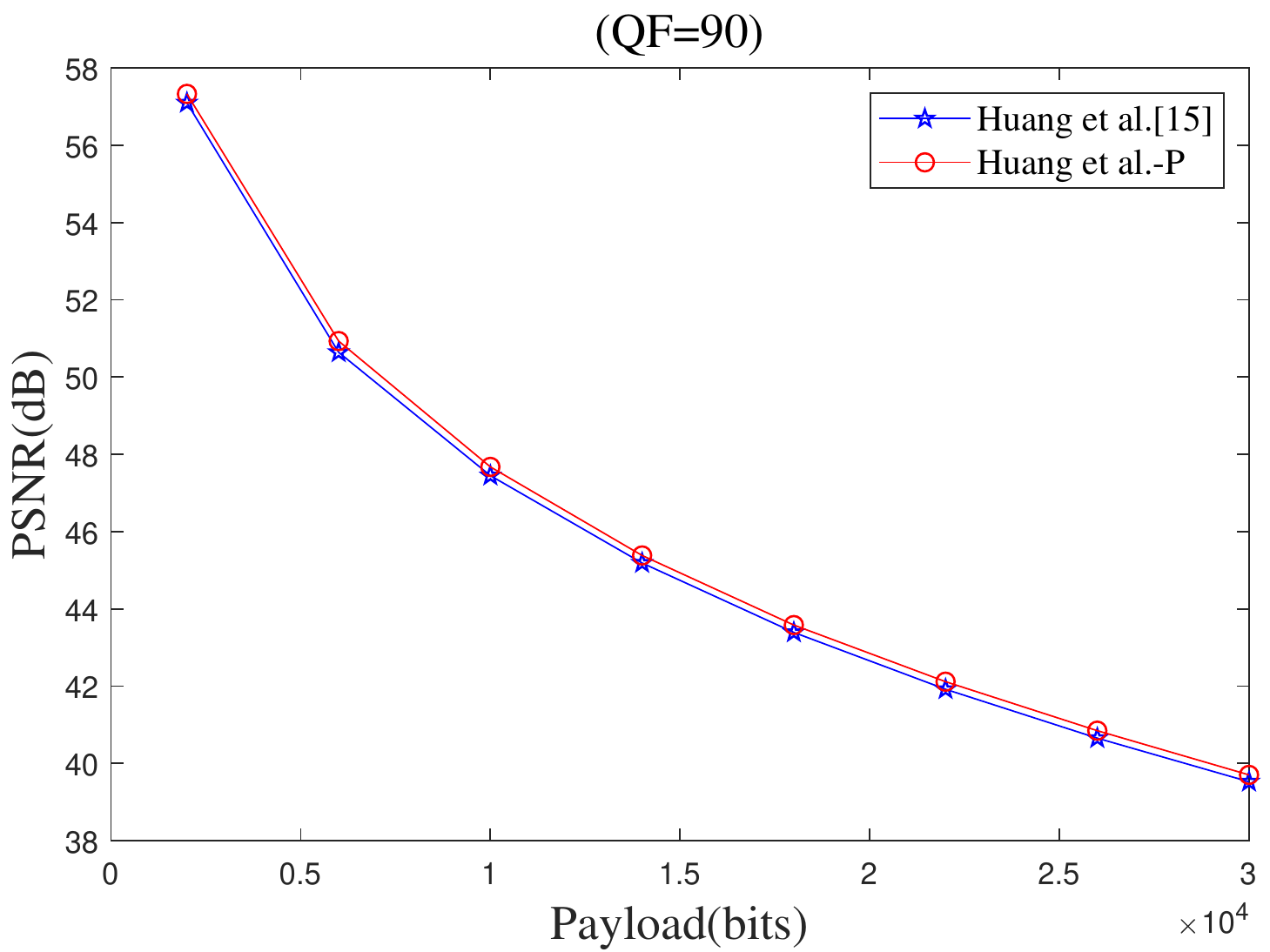}
    \includegraphics[width= 3.1 in]{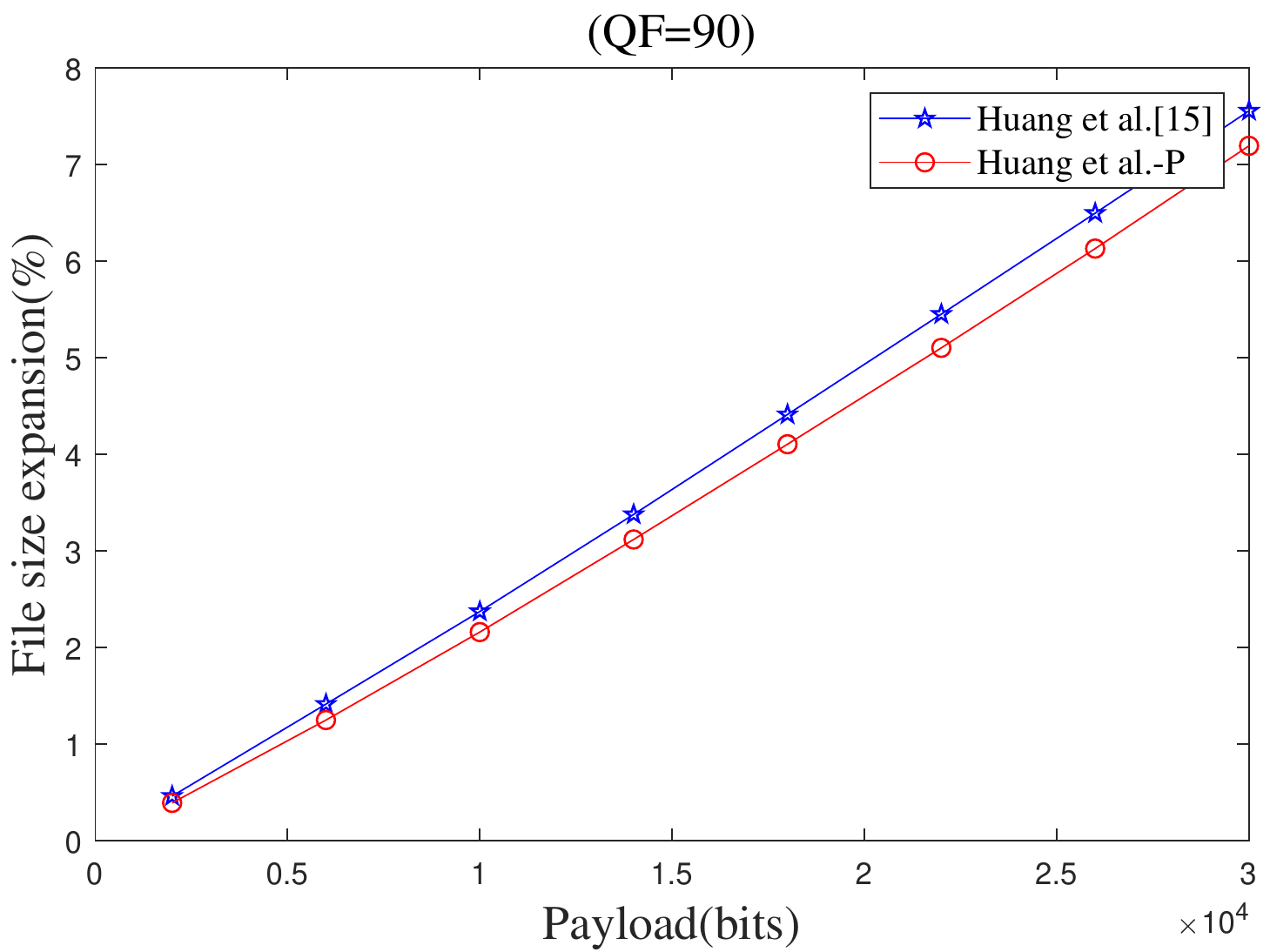}

    }
\caption{Average PSNR and the file size expansion under different payloads between the proposed method and Huang ${et~al. }$'s\cite{IEEEexample:huang2016reversible} method.}
\label{fig8_comparedHuang}
\end{figure*}
\begin{table*}[htbp]
  \centering
  \caption{Comparison of PSNR and increased file size in test images with different QFs and different payloads between the proposed method and Huang ${et~al. }$'s method.}
  \resizebox{380pt}{11.5cm}{
    \begin{tabular}{cccccccccc}
    \toprule
    \multicolumn{2}{c}{\multirow{5}[1]{*}{Lena QF=30}} & \multicolumn{3}{c}{Payload(bits)} & 2000  & 4000  & 6000  & 8000  & 10000 \\
    \multicolumn{2}{c}{} & \multicolumn{2}{c}{\multirow{2}[0]{*}{Huang ${et~al. }$' method\cite{IEEEexample:huang2016reversible}}} & PSNR(dB) & 44.73 & 40.77 & 38.12 & 36.1  & 34.26 \\
    \multicolumn{2}{c}{} & \multicolumn{2}{c}{} & increasement(bits) & 2360  & 5152  & 8160  & 11072 & 14192 \\
    \multicolumn{2}{c}{} & \multicolumn{2}{c}{\multirow{2}[0]{*}{Proposed method}} & PSNR(dB) & 45.28 & 41.63 & 38.8  & 36.52 & 34.41 \\
    \multicolumn{2}{c}{} & \multicolumn{2}{c}{} & increasement(bits) & 2000  & 4272  & 7032  & 10032 & 13696 \\
    \multicolumn{2}{c}{\multirow{5}[0]{*}{Lena QF=50}} & \multicolumn{3}{c}{Payload(bits)} & 6000  & 8000  & 10000 & 12000 & 14000 \\
    \multicolumn{2}{c}{} & \multicolumn{2}{c}{\multirow{2}[0]{*}{Huang ${et~al. }$' method\cite{IEEEexample:huang2016reversible}}} & PSNR(dB) & 42.69  & 40.75  & 38.98  & 37.37  & 35.89  \\
    \multicolumn{2}{c}{} & \multicolumn{2}{c}{} & increasement(bits) & 8248  & 11056 & 14296 & 17376 & 20512 \\
    \multicolumn{2}{c}{} & \multicolumn{2}{c}{\multirow{2}[0]{*}{Proposed method}} & PSNR(dB) & 43.31  & 41.19  & 39.45  & 37.74  & 36.02  \\
    \multicolumn{2}{c}{} & \multicolumn{2}{c}{} & increasement(bits) & 7176  & 10048 & 12920 & 16512 & 20280 \\
    \multicolumn{2}{c}{\multirow{5}[0]{*}{Lena QF=70}} & \multicolumn{3}{c}{Payload(bits)} & 4000  & 8000  & 12000 & 16000 & 20000 \\
    \multicolumn{2}{c}{} & \multicolumn{2}{c}{\multirow{2}[0]{*}{Huang ${et~al. }$' method\cite{IEEEexample:huang2016reversible}}} & PSNR(dB) & 49.21  & 45.36  & 42.42  & 39.84  & 37.25  \\
    \multicolumn{2}{c}{} & \multicolumn{2}{c}{} & increasement(bits) & 5696  & 11176 & 16952 & 23536 & 30520 \\
    \multicolumn{2}{c}{} & \multicolumn{2}{c}{\multirow{2}[0]{*}{Proposed method}} & PSNR(dB) & 49.72  & 45.66  & 42.85  & 40.10  & 37.27  \\
    \multicolumn{2}{c}{} & \multicolumn{2}{c}{} & increasement(bits) & 4616  & 9832  & 15728 & 22248 & 30360 \\
    \multicolumn{2}{c}{\multirow{5}[0]{*}{Lena QF=90}} & \multicolumn{3}{c}{Payload(bits)} & 17000 & 22000 & 27000 & 32000 & 37000 \\
    \multicolumn{2}{c}{} & \multicolumn{2}{c}{\multirow{2}[0]{*}{Huang ${et~al. }$' method\cite{IEEEexample:huang2016reversible}}} & PSNR(dB) & 47.88  & 45.93  & 44.10  & 42.31  & 40.57  \\
    \multicolumn{2}{c}{} & \multicolumn{2}{c}{} & increasement(bits) & 23848 & 31776 & 39976 & 48760 & 57256 \\
    \multicolumn{2}{c}{} & \multicolumn{2}{c}{\multirow{2}[0]{*}{Proposed method}} & PSNR(dB) & 47.97  & 45.98  & 44.19  & 42.40  & 40.64  \\
    \multicolumn{2}{c}{} & \multicolumn{2}{c}{} & increasement(bits) & 22288 & 30264 & 38512 & 47704 & 56728 \\
    \multicolumn{2}{c}{\multirow{5}[0]{*}{Baboon QF=30}} & \multicolumn{3}{c}{Payload(bits)} & 7000  & 12000 & 17000 & 22000 & 27000 \\
    \multicolumn{2}{c}{} & \multicolumn{2}{c}{\multirow{2}[0]{*}{Huang ${et~al. }$' method\cite{IEEEexample:huang2016reversible}}} & PSNR(dB) & 36.79  & 33.06  & 30.57  & 28.75  & 27.25  \\
    \multicolumn{2}{c}{} & \multicolumn{2}{c}{} & increasement(bits) & 10080 & 17120 & 24920 & 32272 & 39704 \\
    \multicolumn{2}{c}{} & \multicolumn{2}{c}{\multirow{2}[0]{*}{Proposed method}} & PSNR(dB) & 37.13  & 33.42  & 30.89  & 29.02  & 27.33  \\
    \multicolumn{2}{c}{} & \multicolumn{2}{c}{} & increasement(bits) & 8576  & 15552 & 22824 & 30784 & 38936 \\
    \multicolumn{2}{c}{\multirow{5}[0]{*}{Baboon QF=50}} & \multicolumn{3}{c}{Payload(bits)} & 12000 & 17000 & 22000 & 27000 & 32000 \\
    \multicolumn{2}{c}{} & \multicolumn{2}{c}{\multirow{2}[0]{*}{Huang ${et~al. }$' method\cite{IEEEexample:huang2016reversible}}} & PSNR(dB) & 36.62  & 33.64  & 31.41  & 29.70  & 28.32  \\
    \multicolumn{2}{c}{} & \multicolumn{2}{c}{} & increasement(bits) & 16920 & 25288 & 33720 & 42600 & 51160 \\
    \multicolumn{2}{c}{} & \multicolumn{2}{c}{\multirow{2}[0]{*}{Proposed method}} & PSNR(dB) & 36.64  & 33.79  & 31.67  & 29.89  & 28.42  \\
    \multicolumn{2}{c}{} & \multicolumn{2}{c}{} & increasement(bits) & 15592 & 23840 & 31800 & 40400 & 49736 \\
    \multicolumn{2}{c}{\multirow{5}[0]{*}{Baboon QF=70}} & \multicolumn{3}{c}{Payload(bits)} & 2000  & 12000 & 22000 & 32000 & 42000 \\
    \multicolumn{2}{c}{} & \multicolumn{2}{c}{\multirow{2}[0]{*}{Huang ${et~al. }$' method\cite{IEEEexample:huang2016reversible}}} & PSNR(dB) & 50.28  & 39.62  & 34.27  & 31.01  & 28.85  \\
    \multicolumn{2}{c}{} & \multicolumn{2}{c}{} & increasement(bits) & 2984  & 17720 & 35664 & 54248 & 69744 \\
    \multicolumn{2}{c}{} & \multicolumn{2}{c}{\multirow{2}[0]{*}{Proposed method}} & PSNR(dB) & 50.53  & 39.73  & 34.37  & 31.14  & 28.87  \\
    \multicolumn{2}{c}{} & \multicolumn{2}{c}{} & increasement(bits) & 2744  & 16704 & 33344 & 51312 & 68912 \\
    \multicolumn{2}{c}{\multirow{5}[0]{*}{Baboon QF=90}} & \multicolumn{3}{c}{Payload(bits)} & 22000 & 32000 & 42000 & 52000 & 62000 \\
    \multicolumn{2}{c}{} & \multicolumn{2}{c}{\multirow{2}[0]{*}{Huang ${et~al. }$' method\cite{IEEEexample:huang2016reversible}}} & PSNR(dB) & 40.02  & 37.22  & 35.18  & 33.56  & 32.24  \\
    \multicolumn{2}{c}{} & \multicolumn{2}{c}{} & increasement(bits) & 37696 & 53864 & 70112 & 85760 & 99904 \\
    \multicolumn{2}{c}{} & \multicolumn{2}{c}{\multirow{2}[0]{*}{Proposed method}} & PSNR(dB) & 40.08  & 37.37  & 35.38  & 33.76  & 32.29  \\
    \multicolumn{2}{c}{} & \multicolumn{2}{c}{} & increasement(bits) & 35400 & 51512 & 67160 & 82344 & 99360 \\
    \multicolumn{2}{c}{\multirow{5}[0]{*}{Airplane QF=30}} & \multicolumn{3}{c}{Payload(bits)} & 2000  & 4000  & 6000  & 8000  & 10000 \\
    \multicolumn{2}{c}{} & \multicolumn{2}{c}{\multirow{2}[0]{*}{Huang ${et~al. }$' method\cite{IEEEexample:huang2016reversible}}} & PSNR(dB) & 45.06  & 39.28  & 36.81  & 35.05  & 33.67  \\
    \multicolumn{2}{c}{} & \multicolumn{2}{c}{} & increasement(bits) & 2360  & 5576  & 8568  & 11640 & 14248 \\
    \multicolumn{2}{c}{} & \multicolumn{2}{c}{\multirow{2}[0]{*}{Proposed method}} & PSNR(dB) & 45.01  & 41.10  & 38.30  & 36.13  & 34.32  \\
    \multicolumn{2}{c}{} & \multicolumn{2}{c}{} & increasement(bits) & 2192  & 4176  & 6944  & 9712  & 13032 \\
    \multicolumn{2}{c}{\multirow{5}[0]{*}{Airplane QF=50}} & \multicolumn{3}{c}{Payload(bits)} & 6000  & 8000  & 10000 & 12000 & 14000 \\
    \multicolumn{2}{c}{} & \multicolumn{2}{c}{\multirow{2}[0]{*}{Huang ${et~al. }$' method\cite{IEEEexample:huang2016reversible}}} & PSNR(dB) & 41.63  & 39.60  & 38.01  & 36.56  & 35.36  \\
    \multicolumn{2}{c}{} & \multicolumn{2}{c}{} & increasement(bits) & 8088  & 11216 & 14096 & 17256 & 20400 \\
    \multicolumn{2}{c}{} & \multicolumn{2}{c}{\multirow{2}[0]{*}{Proposed method}} & PSNR(dB) & 43.12  & 40.86  & 38.91  & 37.23  & 35.69  \\
    \multicolumn{2}{c}{} & \multicolumn{2}{c}{} & increasement(bits) & 6880  & 9576  & 12432 & 16080 & 19456 \\
    \multicolumn{2}{c}{\multirow{5}[0]{*}{Airplane QF=70}} & \multicolumn{3}{c}{Payload(bits)} & 2000  & 6000  & 10000 & 14000 & 18000 \\
    \multicolumn{2}{c}{} & \multicolumn{2}{c}{\multirow{2}[0]{*}{Huang ${et~al. }$' method\cite{IEEEexample:huang2016reversible}}} & PSNR(dB) & 53.24  & 46.89  & 42.74  & 39.93  & 37.51  \\
    \multicolumn{2}{c}{} & \multicolumn{2}{c}{} & increasement(bits) & 2696  & 7584  & 13456 & 19928 & 26736 \\
    \multicolumn{2}{c}{} & \multicolumn{2}{c}{\multirow{2}[0]{*}{Proposed method}} & PSNR(dB) & 53.75  & 47.93  & 44.04  & 40.63  & 37.84  \\
    \multicolumn{2}{c}{} & \multicolumn{2}{c}{} & increasement(bits) & 2072  & 6904  & 12568 & 18992 & 25992 \\
    \multicolumn{2}{c}{\multirow{5}[0]{*}{Airplane QF=90}} & \multicolumn{3}{c}{Payload(bits)} & 12000 & 17000 & 22000 & 27000 & 32000 \\
    \multicolumn{2}{c}{} & \multicolumn{2}{c}{\multirow{2}[0]{*}{Huang ${et~al. }$' method\cite{IEEEexample:huang2016reversible}}} & PSNR(dB) & 50.31  & 47.78  & 45.21  & 42.76  & 40.81  \\
    \multicolumn{2}{c}{} & \multicolumn{2}{c}{} & increasement(bits) & 16208 & 22872 & 31896 & 41448 & 49856 \\
    \multicolumn{2}{c}{} & \multicolumn{2}{c}{\multirow{2}[0]{*}{Proposed method}} & PSNR(dB) & 51.01  & 48.27  & 45.47  & 43.00  & 40.96  \\
    \multicolumn{2}{c}{} & \multicolumn{2}{c}{} & increasement(bits) & 14408 & 22168 & 30232 & 39920 & 48376 \\
    \multicolumn{2}{c}{\multirow{5}[0]{*}{Peppers QF=30}} & \multicolumn{3}{c}{Payload(bits)} & 5000  & 6000  & 7000  & 8000  & 9000 \\
    \multicolumn{2}{c}{} & \multicolumn{2}{c}{\multirow{2}[0]{*}{Huang ${et~al. }$' method\cite{IEEEexample:huang2016reversible}}} & PSNR(dB) & 39.16  & 37.81  & 36.68  & 35.77  & 34.92  \\
    \multicolumn{2}{c}{} & \multicolumn{2}{c}{} & increasement(bits) & 6472  & 8096  & 9696  & 11192 & 12624 \\
    \multicolumn{2}{c}{} & \multicolumn{2}{c}{\multirow{2}[0]{*}{Proposed method}} & PSNR(dB) & 40.30  & 38.98  & 37.79  & 36.60  & 35.51  \\
    \multicolumn{2}{c}{} & \multicolumn{2}{c}{} & increasement(bits) & 5488  & 7000  & 8440  & 9992  & 12184 \\
    \multicolumn{2}{c}{\multirow{5}[0]{*}{Peppers QF=50}} & \multicolumn{3}{c}{Payload(bits)} & 6000  & 8000  & 10000 & 12000 & 14000 \\
    \multicolumn{2}{c}{} & \multicolumn{2}{c}{\multirow{2}[0]{*}{Huang ${et~al. }$' method\cite{IEEEexample:huang2016reversible}}} & PSNR(dB) & 43.08  & 41.04  & 39.21  & 37.68  & 36.30  \\
    \multicolumn{2}{c}{} & \multicolumn{2}{c}{} & increasement(bits) & 8408  & 11192 & 14232 & 17528 & 20448 \\
    \multicolumn{2}{c}{} & \multicolumn{2}{c}{\multirow{2}[0]{*}{Proposed method}} & PSNR(dB) & 43.73  & 41.75  & 40.02  & 38.33  & 36.68  \\
    \multicolumn{2}{c}{} & \multicolumn{2}{c}{} & increasement(bits) & 6992  & 9920  & 12920 & 16392 & 19816 \\
    \multicolumn{2}{c}{\multirow{5}[0]{*}{Peppers QF=70}} & \multicolumn{3}{c}{Payload(bits)} & 4000  & 8000  & 12000 & 16000 & 20000 \\
    \multicolumn{2}{c}{} & \multicolumn{2}{c}{\multirow{2}[0]{*}{Huang ${et~al. }$' method\cite{IEEEexample:huang2016reversible}}} & PSNR(dB) & 49.09  & 45.53  & 42.84  & 40.73  & 38.47  \\
    \multicolumn{2}{c}{} & \multicolumn{2}{c}{} & increasement(bits) & 6040  & 11656 & 17056 & 23048 & 29320 \\
    \multicolumn{2}{c}{} & \multicolumn{2}{c}{\multirow{2}[0]{*}{Proposed method}} & PSNR(dB) & 49.71  & 46.01  & 43.55  & 41.26  & 38.74  \\
    \multicolumn{2}{c}{} & \multicolumn{2}{c}{} & increasement(bits) & 4824  & 9736  & 15400 & 21760 & 28160 \\
    \multicolumn{2}{c}{\multirow{5}[1]{*}{Peppers QF=90}} & \multicolumn{3}{c}{Payload(bits)} & 2000  & 12000 & 22000 & 32000 & 42000 \\
    \multicolumn{2}{c}{} & \multicolumn{2}{c}{\multirow{2}[0]{*}{Huang ${et~al. }$' method\cite{IEEEexample:huang2016reversible}}} & PSNR(dB) & 57.81  & 48.19  & 44.29  & 41.59  & 39.28  \\
    \multicolumn{2}{c}{} & \multicolumn{2}{c}{} & increasement(bits) & 2720  & 17576 & 33120 & 50128 & 67536 \\
    \multicolumn{2}{c}{} & \multicolumn{2}{c}{\multirow{2}[1]{*}{Proposed method}} & PSNR(dB) & 57.46  & 48.30  & 44.40  & 41.67  & 39.33  \\
    \multicolumn{2}{c}{} & \multicolumn{2}{c}{} & increasement(bits) & 2688  & 16544 & 32224 & 48904 & 66368 \\
    \bottomrule
    \end{tabular}%
    }
  \label{tab3_comparedHuang}%
\end{table*}%
\par The comparisons with \cite{IEEEexample:huang2016reversible} and \cite{IEEEexample:hou2018reversible} are both based on the average results of 96 images for better contrast. Farther, the results of the four test images are also shown in Table~\ref{tab3_comparedHuang} and Table~\ref{tab4_comparedHou} to clearly see the improvement. The reason why the proposed method compares with the other two schemes individually is that the proposed method is carried on the basis of each scheme.
\subsection{Comparison with Huang ${ et~al. }$'s scheme}
\label{Comparison with Huang scheme}
\par As we can see from the process of Huang ${ et~al. }$'s scheme, the operation is done in units of 8${\times}$8 sized DCT block. Therefore, we combine the multi-objective optimization with Huang ${ et~al. }$'s method on the basis of taking the set of 8${\times}$8 sized DCT blocks as cover in our proposed scheme.
\par In this part, the comparison with Huang ${ et~al. }$'s scheme is shown as Fig.~\ref{fig8_comparedHuang}. All the results are gained by the average values of 96 images. And the x-axis represents the payload, the y-axis represents the Peak Signal to Noise Ratio (PSNR) and the file size expansion respectively in the left and the right column. The PSNR can reflect the visual quality of the stego cover, the higher it represents the smaller the image distortion caused in the embedding process. In addition, the file size expansion expresses the increas of the file size after embedding data, which is expected very small. And the proposed scheme is named Huang ${ et~al. }$-P in the experimental results. We can obverse that the results of our proposed scheme is higher than results of Huang ${ et~al. }$'s scheme in the left graph and the results of our proposed scheme is lower than results of Huang ${ et~al. }$'s scheme in the right and in the left, the higher line means the good image quality of the stego image and in the right, the lower line indicates the small file size expansion. So, it is obvious that the proposed scheme does better than the scheme of Huang ${et~al. }$ both in the image quality and file size expansion.
\par What's more, the improvement of our proposed multi-objective optimization is getting smaller as the QF of JPEG images increases. The closer result shows that the advantage of the multi-objective optimization is weakening in the low compression image. It is because that the bigger QF means the redundant space of the image is more and.
\par Furthermore, the specific values of the four test images with four different QFs are shown in Table \ref{tab3_comparedHuang}. In the case of each QF for each image, the first line is the embedding capacity, the second and third lines are respectively the PSNR and increased file size of the stego cover generated by Huang ${ et~al. }$'s method, and the fourth and fifth lines are the PSNR and file size expansion of our generated stego cover.
\begin{figure*}
\centering{
    \includegraphics[width= 3.1 in]{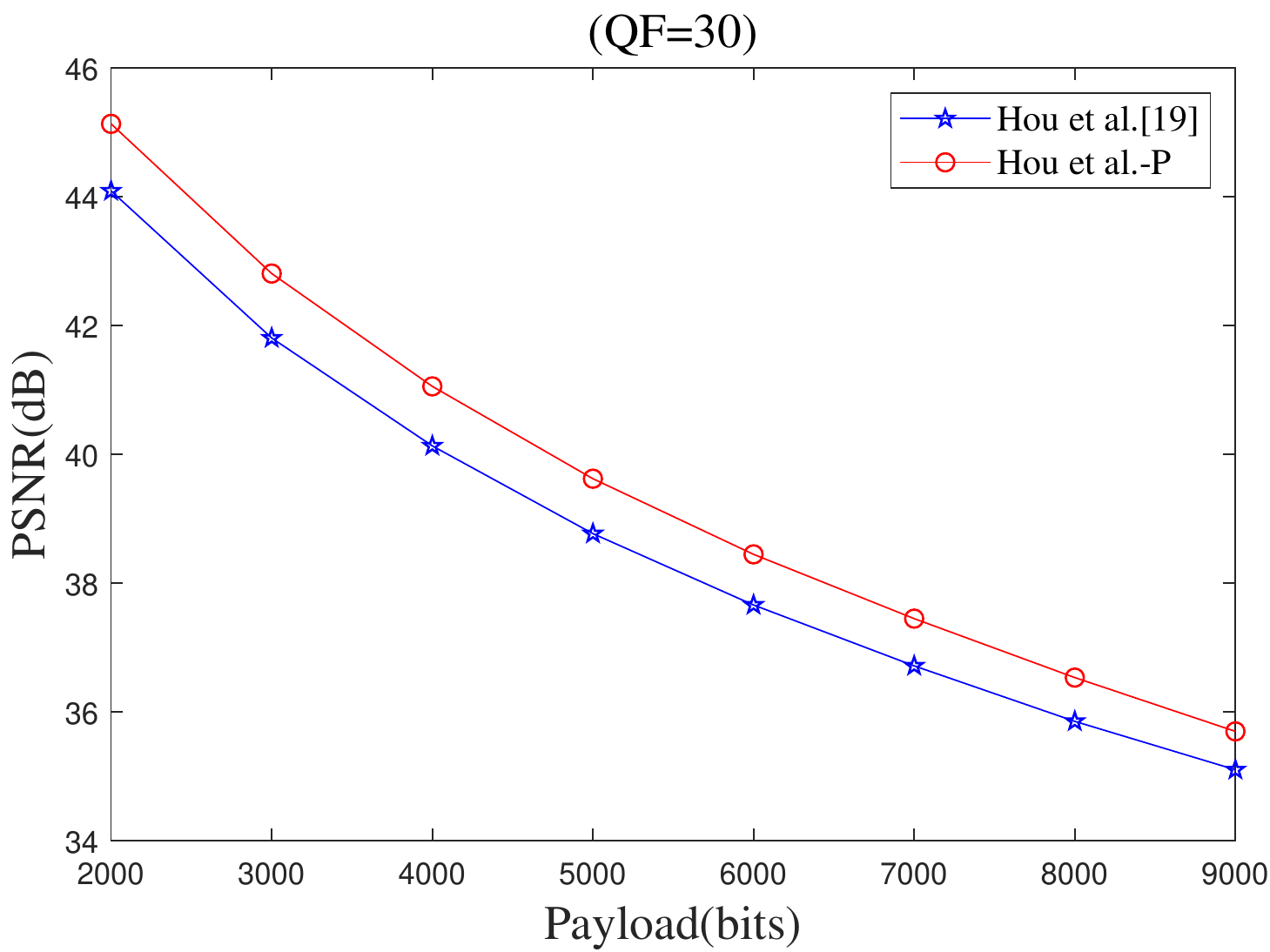}
    \includegraphics[width= 3.1 in]{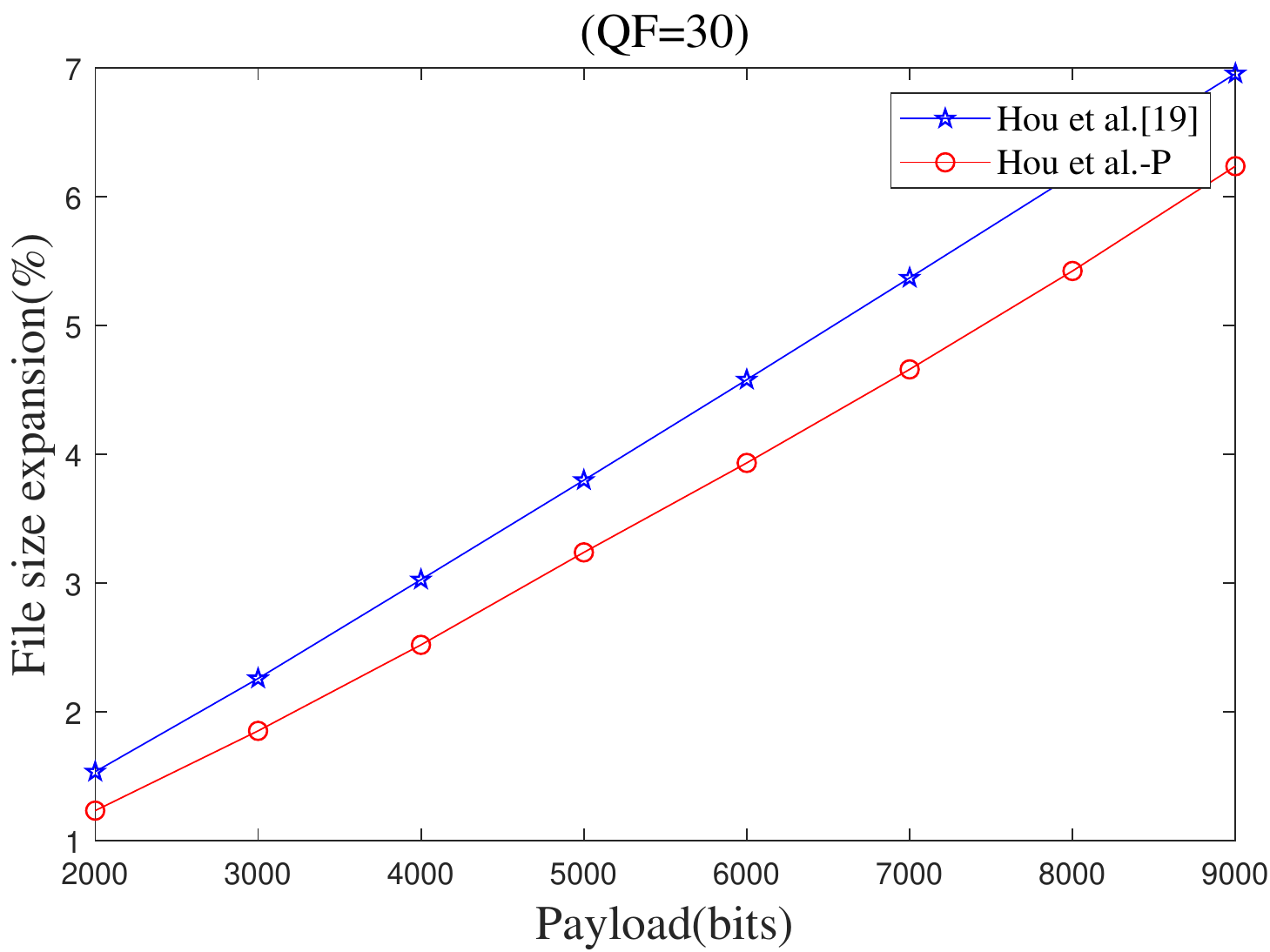}
    \includegraphics[width= 3.1 in]{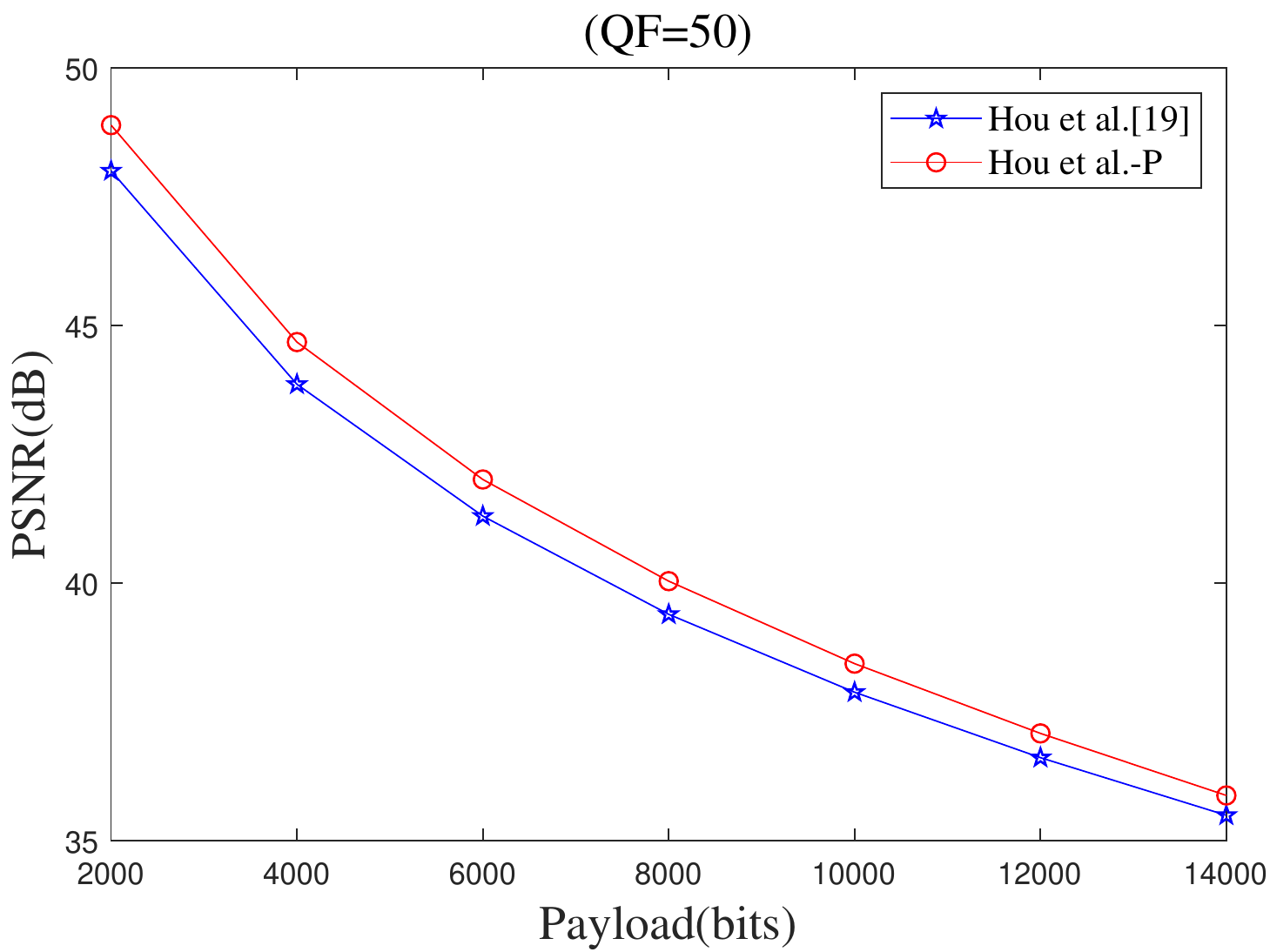}
    \includegraphics[width= 3.1 in]{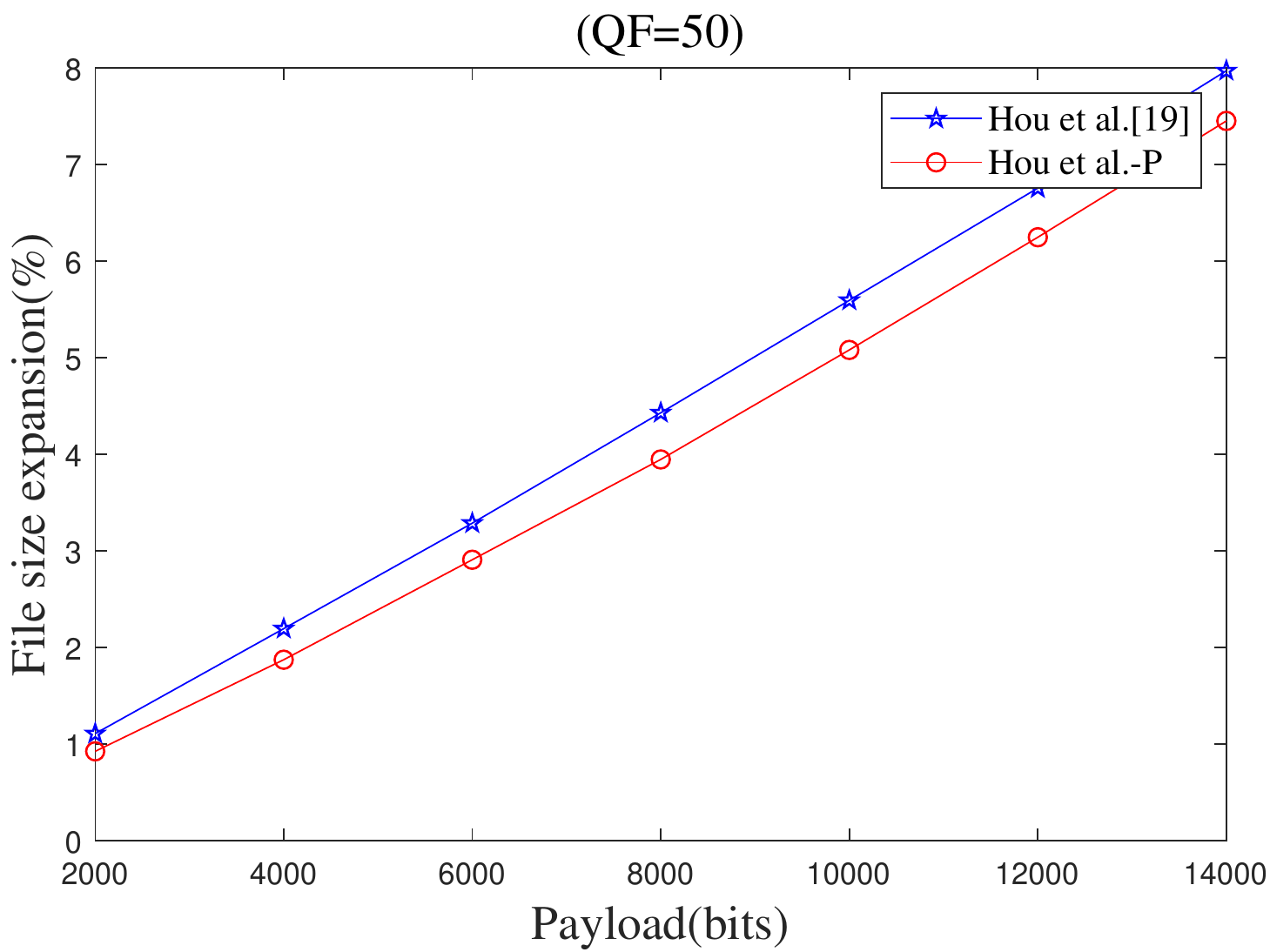}
    \includegraphics[width= 3.1 in]{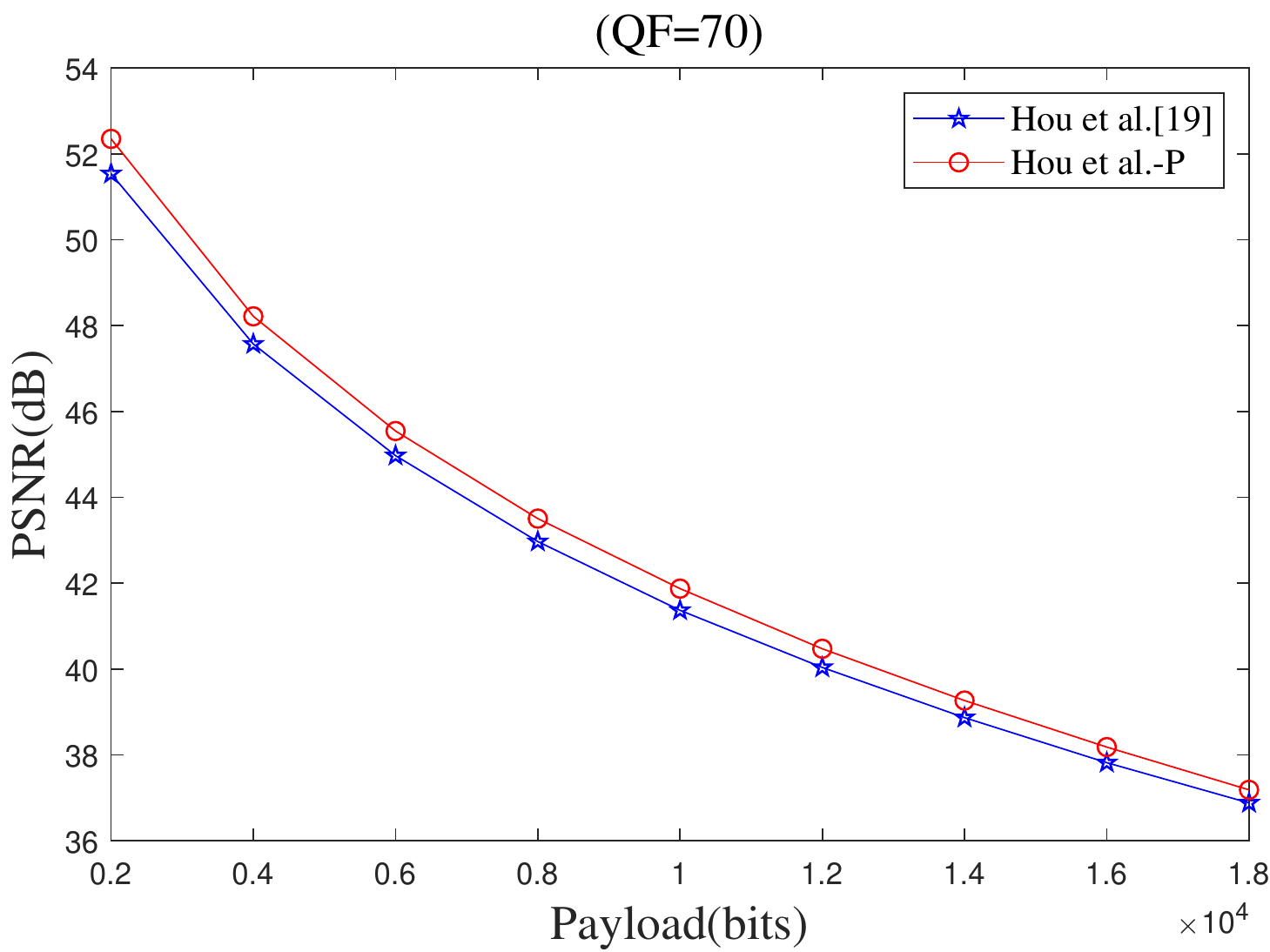}
    \includegraphics[width= 3.1 in]{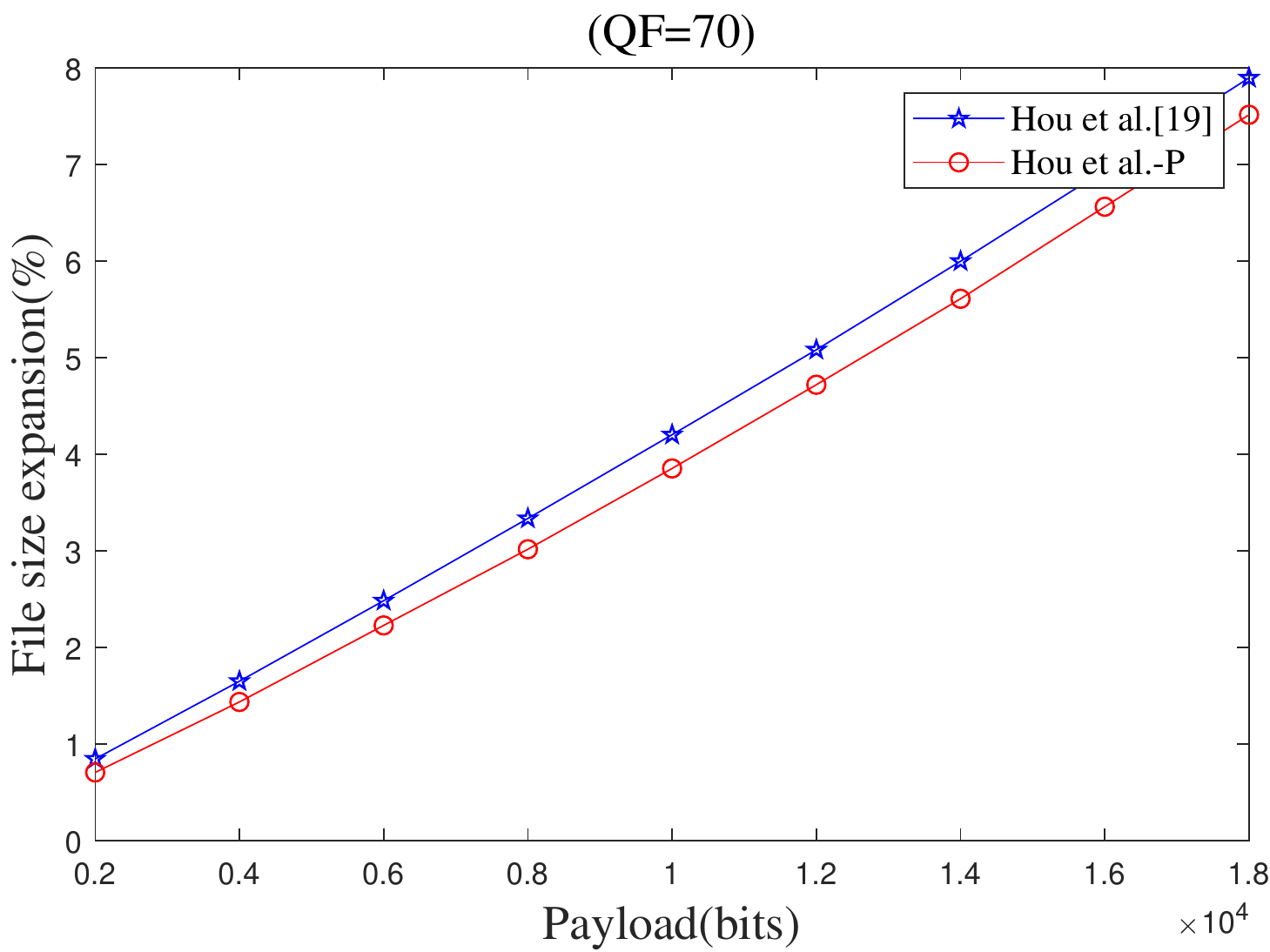}
    \includegraphics[width= 3.1 in]{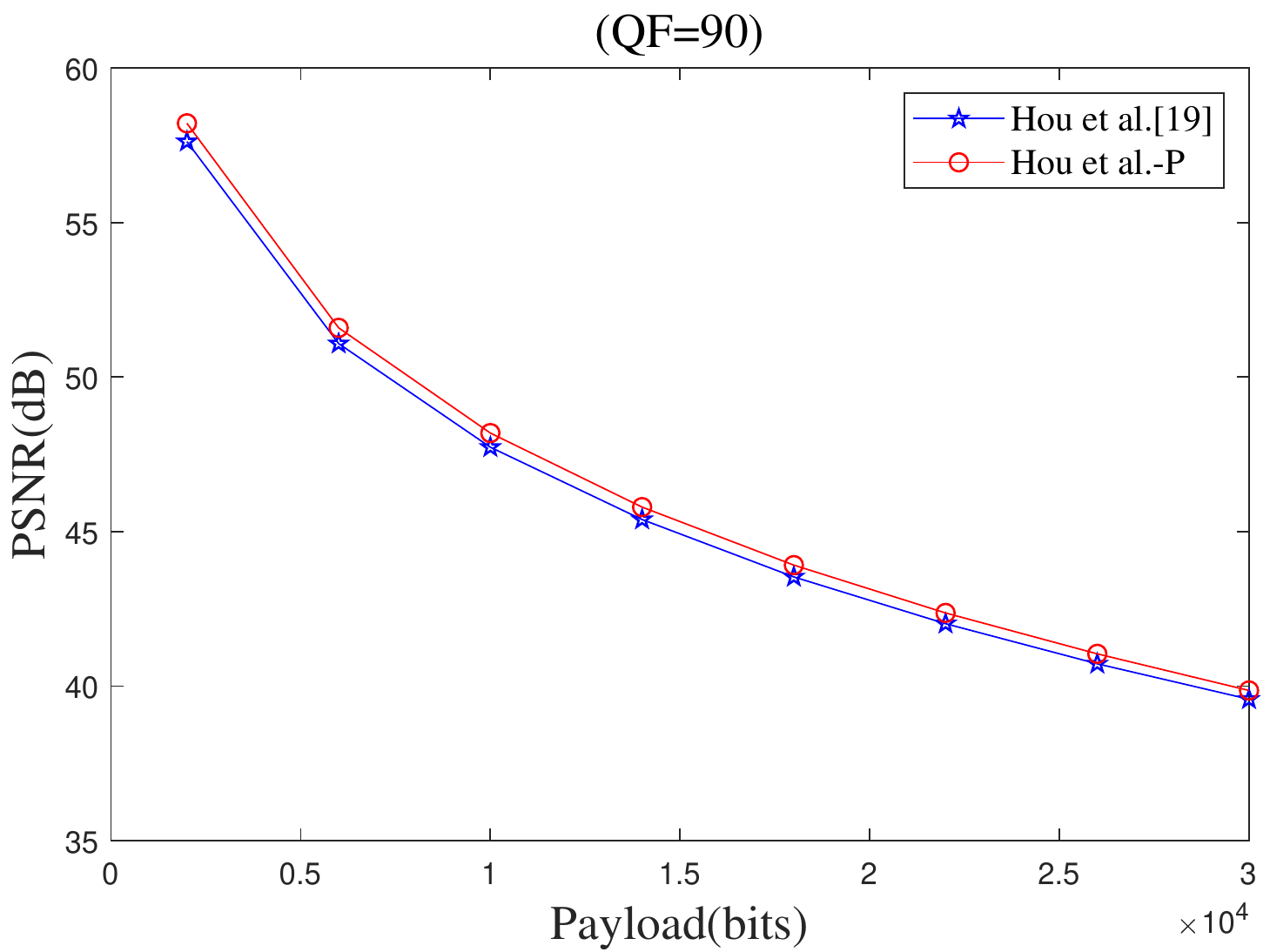}
    \includegraphics[width= 3.1 in]{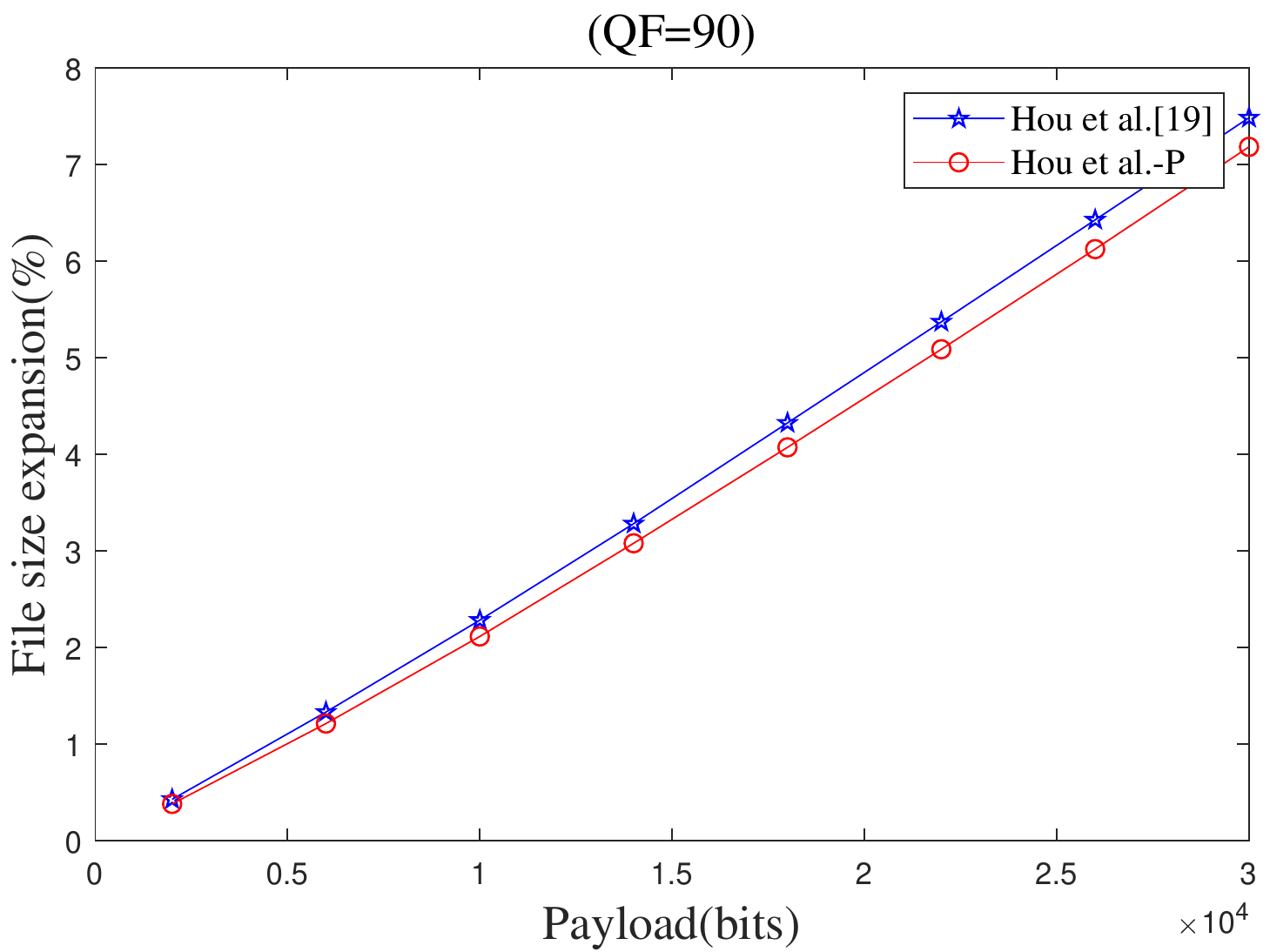}
    }
\caption{Average PSNR and the file size expansion under different payloads between the proposed method and Hou ${et~al. }$'s\cite{IEEEexample:hou2018reversible} method.}
\label{fig9_comparedHou}
\end{figure*}

\begin{table*}[htbp]
  \centering
  \caption{Comparison of PSNR and increased file size in test images with different QFs and different payloads between the proposed method and Hou ${et~al. }$'s method.}
  \resizebox{380pt}{11cm}{
    \begin{tabular}{cccccccccc}
    \toprule
    \multicolumn{2}{c}{\multirow{5}[1]{*}{Lena QF=30}} & \multicolumn{3}{c}{Payload(bits)} & 2000  & 4000  & 6000  & 8000  & 10000 \\
    \multicolumn{2}{c}{} & \multicolumn{2}{c}{\multirow{2}[0]{*}{Hou ${et~al. }$' method\cite{IEEEexample:hou2018reversible}}} & PSNR(dB) & 44.78  & 40.76  & 38.29  & 36.14  & 34.35  \\
    \multicolumn{2}{c}{} & \multicolumn{2}{c}{} & increasement(bits) & 2848  & 5200  & 7992  & 11184 & 14016 \\
    \multicolumn{2}{c}{} & \multicolumn{2}{c}{\multirow{2}[0]{*}{Proposed method}} & PSNR(dB) & 45.94  & 42.14  & 39.09  & 36.68  & 34.50  \\
    \multicolumn{2}{c}{} & \multicolumn{2}{c}{} & increasement(bits) & 2352  & 4064  & 6776  & 10008 & 13808 \\
    \multicolumn{2}{c}{\multirow{5}[0]{*}{Lena QF=50}} & \multicolumn{3}{c}{Payload(bits)} & 6000  & 8000  & 10000 & 12000 & 14000 \\
    \multicolumn{2}{c}{} & \multicolumn{2}{c}{\multirow{2}[0]{*}{Hou ${et~al. }$' method\cite{IEEEexample:hou2018reversible}}} & PSNR(dB) & 42.83  & 40.97  & 39.14  & 37.53  & 35.97  \\
    \multicolumn{2}{c}{} & \multicolumn{2}{c}{} & increasement(bits) & 8072  & 10976 & 14096 & 17520 & 20368 \\
    \multicolumn{2}{c}{} & \multicolumn{2}{c}{\multirow{2}[0]{*}{Proposed method}} & PSNR(dB) & 43.64  & 41.55  & 39.57  & 37.90  & 36.00  \\
    \multicolumn{2}{c}{} & \multicolumn{2}{c}{} & increasement(bits) & 7328  & 10072 & 13064 & 16152 & 20480 \\
    \multicolumn{2}{c}{\multirow{5}[0]{*}{Lena QF=70}} & \multicolumn{3}{c}{Payload(bits)} & 4000  & 8000  & 12000 & 16000 & 20000 \\
    \multicolumn{2}{c}{} & \multicolumn{2}{c}{\multirow{2}[0]{*}{Hou ${et~al. }$' method\cite{IEEEexample:hou2018reversible}}} & PSNR(dB) & 49.85  & 45.54  & 42.69  & 39.95  & 37.28  \\
    \multicolumn{2}{c}{} & \multicolumn{2}{c}{} & increasement(bits) & 5504  & 11280 & 16768 & 23160 & 30360 \\
    \multicolumn{2}{c}{} & \multicolumn{2}{c}{\multirow{2}[0]{*}{Proposed method}} & PSNR(dB) & 50.47  & 45.99  & 42.94  & 40.33  & 37.39  \\
    \multicolumn{2}{c}{} & \multicolumn{2}{c}{} & increasement(bits) & 4968  & 10416 & 16384 & 22448 & 30288 \\
    \multicolumn{2}{c}{\multirow{5}[0]{*}{Lena QF=90}} & \multicolumn{3}{c}{Payload(bits)} & 17000 & 22000 & 27000 & 32000 & 37000 \\
    \multicolumn{2}{c}{} & \multicolumn{2}{c}{\multirow{2}[0]{*}{Hou ${et~al. }$' method\cite{IEEEexample:hou2018reversible}}} & PSNR(dB) & 47.92  & 45.95  & 44.13  & 42.36  & 40.61  \\
    \multicolumn{2}{c}{} & \multicolumn{2}{c}{} & increasement(bits) & 23520 & 31432 & 39960 & 48816 & 57000 \\
    \multicolumn{2}{c}{} & \multicolumn{2}{c}{\multirow{2}[0]{*}{Proposed method}} & PSNR(dB) & 48.25  & 46.29  & 44.33  & 42.54  & 40.69  \\
    \multicolumn{2}{c}{} & \multicolumn{2}{c}{} & increasement(bits) & 22536 & 30248 & 38784 & 47552 & 56912 \\
    \multicolumn{2}{c}{\multirow{5}[0]{*}{Baboon QF=30}} & \multicolumn{3}{c}{Payload(bits)} & 7000  & 12000 & 17000 & 22000 & 27000 \\
    \multicolumn{2}{c}{} & \multicolumn{2}{c}{\multirow{2}[0]{*}{Hou ${et~al. }$' method\cite{IEEEexample:hou2018reversible}}} & PSNR(dB) & 37.13  & 33.47  & 30.96  & 28.96  & 27.28  \\
    \multicolumn{2}{c}{} & \multicolumn{2}{c}{} & increasement(bits) & 9616  & 16536 & 24144 & 31432 & 39336 \\
    \multicolumn{2}{c}{} & \multicolumn{2}{c}{\multirow{2}[0]{*}{Proposed method}} & PSNR(dB) & 37.66  & 33.95  & 31.30  & 29.26  & 27.39  \\
    \multicolumn{2}{c}{} & \multicolumn{2}{c}{} & increasement(bits) & 8864  & 15368 & 22664 & 29744 & 38496 \\
    \multicolumn{2}{c}{\multirow{5}[0]{*}{Baboon QF=50}} & \multicolumn{3}{c}{Payload(bits)} & 12000 & 17000 & 22000 & 27000 & 32000 \\
    \multicolumn{2}{c}{} & \multicolumn{2}{c}{\multirow{2}[0]{*}{Hou ${et~al. }$' method\cite{IEEEexample:hou2018reversible}}} & PSNR(dB) & 36.90  & 33.95  & 31.76  & 29.91  & 28.50  \\
    \multicolumn{2}{c}{} & \multicolumn{2}{c}{} & increasement(bits) & 16712 & 24544 & 32488 & 42368 & 50256 \\
    \multicolumn{2}{c}{} & \multicolumn{2}{c}{\multirow{2}[0]{*}{Proposed method}} & PSNR(dB) & 37.25  & 34.24  & 31.97  & 30.08  & 28.60  \\
    \multicolumn{2}{c}{} & \multicolumn{2}{c}{} & increasement(bits) & 15512 & 23320 & 31904 & 40704 & 49088 \\
    \multicolumn{2}{c}{\multirow{5}[0]{*}{Baboon QF=70}} & \multicolumn{3}{c}{Payload(bits)} & 2000  & 12000 & 22000 & 32000 & 42000 \\
    \multicolumn{2}{c}{} & \multicolumn{2}{c}{\multirow{2}[0]{*}{Hou ${et~al. }$' method\cite{IEEEexample:hou2018reversible}}} & PSNR(dB) & 50.77  & 39.85  & 34.46  & 31.10  & 28.86  \\
    \multicolumn{2}{c}{} & \multicolumn{2}{c}{} & increasement(bits) & 2856  & 17640 & 35296 & 53400 & 69864 \\
    \multicolumn{2}{c}{} & \multicolumn{2}{c}{\multirow{2}[0]{*}{Proposed method}} & PSNR(dB) & 51.24  & 40.14  & 34.74  & 31.39  & 28.92  \\
    \multicolumn{2}{c}{} & \multicolumn{2}{c}{} & increasement(bits) & 2624  & 16728 & 34080 & 50840 & 68760 \\
    \multicolumn{2}{c}{\multirow{5}[0]{*}{Baboon QF=90}} & \multicolumn{3}{c}{Payload(bits)} & 22000 & 32000 & 42000 & 52000 & 62000 \\
    \multicolumn{2}{c}{} & \multicolumn{2}{c}{\multirow{2}[0]{*}{Hou ${et~al. }$' method\cite{IEEEexample:hou2018reversible}}} & PSNR(dB) & 40.03  & 37.26  & 35.19  & 33.60  & 32.25  \\
    \multicolumn{2}{c}{} & \multicolumn{2}{c}{} & increasement(bits) & 37264 & 54192 & 69808 & 85336 & 100168 \\
    \multicolumn{2}{c}{} & \multicolumn{2}{c}{\multirow{2}[0]{*}{Proposed method}} & PSNR(dB) & 40.28  & 37.52  & 35.48  & 33.81  & 32.29  \\
    \multicolumn{2}{c}{} & \multicolumn{2}{c}{} & increasement(bits) & 36184 & 51688 & 67392 & 82864 & 99552 \\
    \multicolumn{2}{c}{\multirow{5}[0]{*}{Airplane QF=30}} & \multicolumn{3}{c}{Payload(bits)} & 2000  & 4000  & 6000  & 8000  & 10000 \\
    \multicolumn{2}{c}{} & \multicolumn{2}{c}{\multirow{2}[0]{*}{Hou ${et~al. }$' method\cite{IEEEexample:hou2018reversible}}} & PSNR(dB) & 45.30  & 40.59  & 37.81  & 35.67  & 33.88  \\
    \multicolumn{2}{c}{} & \multicolumn{2}{c}{} & increasement(bits) & 2232  & 5128  & 7904  & 11168 & 14168 \\
    \multicolumn{2}{c}{} & \multicolumn{2}{c}{\multirow{2}[0]{*}{Proposed method}} & PSNR(dB) & 45.95  & 41.53  & 38.61  & 36.27  & 34.37  \\
    \multicolumn{2}{c}{} & \multicolumn{2}{c}{} & increasement(bits) & 1952  & 4344  & 6744  & 9832  & 13208 \\
    \multicolumn{2}{c}{\multirow{5}[0]{*}{Airplane QF=50}} & \multicolumn{3}{c}{Payload(bits)} & 6000  & 8000  & 10000 & 12000 & 14000 \\
    \multicolumn{2}{c}{} & \multicolumn{2}{c}{\multirow{2}[0]{*}{Hou ${et~al. }$' method\cite{IEEEexample:hou2018reversible}}} & PSNR(dB) & 42.74  & 40.50  & 38.37  & 36.89  & 35.37  \\
    \multicolumn{2}{c}{} & \multicolumn{2}{c}{} & increasement(bits) & 7744  & 10448 & 14048 & 16896 & 20536 \\
    \multicolumn{2}{c}{} & \multicolumn{2}{c}{\multirow{2}[0]{*}{Proposed method}} & PSNR(dB) & 43.41  & 41.08  & 39.09  & 37.33  & 35.74  \\
    \multicolumn{2}{c}{} & \multicolumn{2}{c}{} & increasement(bits) & 6864  & 9752  & 12760 & 16040 & 19360 \\
    \multicolumn{2}{c}{\multirow{5}[0]{*}{Airplane QF=70}} & \multicolumn{3}{c}{Payload(bits)} & 2000  & 6000  & 10000 & 14000 & 18000 \\
    \multicolumn{2}{c}{} & \multicolumn{2}{c}{\multirow{2}[0]{*}{Hou ${et~al. }$' method\cite{IEEEexample:hou2018reversible}}} & PSNR(dB) & 53.81  & 47.81  & 43.80  & 40.40  & 37.60  \\
    \multicolumn{2}{c}{} & \multicolumn{2}{c}{} & increasement(bits) & 2928  & 7352  & 13376 & 20008 & 26496 \\
    \multicolumn{2}{c}{} & \multicolumn{2}{c}{\multirow{2}[0]{*}{Proposed method}} & PSNR(dB) & 54.48  & 48.23  & 44.23  & 40.76  & 37.96  \\
    \multicolumn{2}{c}{} & \multicolumn{2}{c}{} & increasement(bits) & 2520  & 7072  & 12544 & 18960 & 25744 \\
    \multicolumn{2}{c}{\multirow{5}[0]{*}{Airplane QF=90}} & \multicolumn{3}{c}{Payload(bits)} & 12000 & 17000 & 22000 & 27000 & 32000 \\
    \multicolumn{2}{c}{} & \multicolumn{2}{c}{\multirow{2}[0]{*}{Hou ${et~al. }$' method\cite{IEEEexample:hou2018reversible}}} & PSNR(dB) & 50.84  & 47.98  & 45.32  & 42.79  & 40.83  \\
    \multicolumn{2}{c}{} & \multicolumn{2}{c}{} & increasement(bits) & 15400 & 22872 & 31816 & 41040 & 49048 \\
    \multicolumn{2}{c}{} & \multicolumn{2}{c}{\multirow{2}[0]{*}{Proposed method}} & PSNR(dB) & 51.15  & 48.45  & 45.52  & 43.13  & 40.94  \\
    \multicolumn{2}{c}{} & \multicolumn{2}{c}{} & increasement(bits) & 14888 & 22080 & 30800 & 39688 & 49368 \\
    \multicolumn{2}{c}{\multirow{5}[0]{*}{Peppers QF=30}} & \multicolumn{3}{c}{Payload(bits)} & 5000  & 6000  & 7000  & 8000  & 9000 \\
    \multicolumn{2}{c}{} & \multicolumn{2}{c}{\multirow{2}[0]{*}{Hou ${et~al. }$' method\cite{IEEEexample:hou2018reversible}}} & PSNR(dB) & 39.68  & 38.45  & 37.29  & 36.25  & 35.22  \\
    \multicolumn{2}{c}{} & \multicolumn{2}{c}{} & increasement(bits) & 6192  & 7528  & 9648  & 10824 & 12768 \\
    \multicolumn{2}{c}{} & \multicolumn{2}{c}{\multirow{2}[0]{*}{Proposed method}} & PSNR(dB) & 40.57  & 39.22  & 37.90  & 36.73  & 35.54  \\
    \multicolumn{2}{c}{} & \multicolumn{2}{c}{} & increasement(bits) & 5480  & 6768  & 8544  & 10440 & 12128 \\
    \multicolumn{2}{c}{\multirow{5}[0]{*}{Peppers QF=50}} & \multicolumn{3}{c}{Payload(bits)} & 6000  & 8000  & 10000 & 12000 & 14000 \\
    \multicolumn{2}{c}{} & \multicolumn{2}{c}{\multirow{2}[0]{*}{Hou ${et~al. }$' method\cite{IEEEexample:hou2018reversible}}} & PSNR(dB) & 43.34  & 41.48  & 39.83  & 38.19  & 36.45  \\
    \multicolumn{2}{c}{} & \multicolumn{2}{c}{} & increasement(bits) & 8304  & 10840 & 13888 & 17072 & 20672 \\
    \multicolumn{2}{c}{} & \multicolumn{2}{c}{\multirow{2}[0]{*}{Proposed method}} & PSNR(dB) & 44.04  & 42.12  & 40.25  & 38.52  & 36.74  \\
    \multicolumn{2}{c}{} & \multicolumn{2}{c}{} & increasement(bits) & 7264  & 9664  & 12968 & 16328 & 19928 \\
    \multicolumn{2}{c}{\multirow{5}[0]{*}{Peppers QF=70}} & \multicolumn{3}{c}{Payload(bits)} & 4000  & 8000  & 12000 & 16000 & 20000 \\
    \multicolumn{2}{c}{} & \multicolumn{2}{c}{\multirow{2}[0]{*}{Hou ${et~al. }$' method\cite{IEEEexample:hou2018reversible}}} & PSNR(dB) & 49.89  & 46.00  & 43.40  & 41.16  & 38.56  \\
    \multicolumn{2}{c}{} & \multicolumn{2}{c}{} & increasement(bits) & 5616  & 11544 & 16640 & 22544 & 28944 \\
    \multicolumn{2}{c}{} & \multicolumn{2}{c}{\multirow{2}[0]{*}{Proposed method}} & PSNR(dB) & 50.60  & 46.62  & 43.75  & 41.40  & 38.85  \\
    \multicolumn{2}{c}{} & \multicolumn{2}{c}{} & increasement(bits) & 4488  & 9856  & 15712 & 21856 & 28280 \\
    \multicolumn{2}{c}{\multirow{5}[1]{*}{Peppers QF=90}} & \multicolumn{3}{c}{Payload(bits)} & 2000  & 12000 & 22000 & 32000 & 42000 \\
    \multicolumn{2}{c}{} & \multicolumn{2}{c}{\multirow{2}[0]{*}{Hou ${et~al. }$' method\cite{IEEEexample:hou2018reversible}}} & PSNR(dB) & 59.02  & 48.82  & 44.54  & 41.64  & 39.18  \\
    \multicolumn{2}{c}{} & \multicolumn{2}{c}{} & increasement(bits) & 2248  & 16920 & 32256 & 50320 & 68488 \\
    \multicolumn{2}{c}{} & \multicolumn{2}{c}{\multirow{2}[1]{*}{Proposed method}} & PSNR(dB) & 59.52  & 49.21  & 44.83  & 41.83  & 39.35  \\
    \multicolumn{2}{c}{} & \multicolumn{2}{c}{} & increasement(bits) & 2672  & 15752 & 31320 & 48520 & 66528 \\
    \bottomrule
    \end{tabular}%
    }
  \label{tab4_comparedHou}%
\end{table*}%
\subsection{Comparison with Hou ${ et~al. }$'s scheme}
\par It can be seen from the analysis in section \ref{Related works} that the cover of Hou ${ et~al. }$'s scheme is 8${\times}$8 sized blocks which are different from original 8${\times}$8 sized DCT blocks. The cover is a set of blocks that choose values in some frequencies from the original blocks and set values in the other frequencies as 0. The multi-objective optimization is performed combined with the Hou ${ et~al. }$'s scheme in this experiment to show the improvement of the proposed scheme in Fig.~\ref{fig9_comparedHou}.
\par The meanings of x-axis and y-axis are the same as the description in section~\ref{Comparison with Huang scheme}. And the result of our experiment is marker as Hou ${ et~al. }$-P in this experiment. What's more, not only the image quality gained by our scheme is higher than that of Hou ${ et~al. }$, which is obvious in the left graph in Fig.~\ref{fig9_comparedHou} that the red line is higher than the blue one as the payload increases, but also the file size expansion of our method is lower than the expansion of Hou ${ et~al. }$, expressed as the red line lower than the blue line in the right graph in Fig.~\ref{fig9_comparedHou}. In addition, the specific values of the four test images compared with the result of Hou ${ et~al. }$'s method are shown in Table \ref{tab4_comparedHou}.
And the meaning of each line in the table is almost the same as that in Table~\ref{tab3_comparedHuang}, expect that the second and third lines under each QF for each image are respectively the PSNR and increased file size of the cover generated by Hou ${ et~al. }$'s method.
\section{Conclusion}
\par A novel RDH scheme in JPEG images is proposed in this paper which applies multi-objective optimization. Most state-of-the-art RDH schemes in JPEG images only consider the image quality and ignore the file size expansion in designing methods. However, the file size is an important aspect for JPEG images, so multi-objective optimization is used in our proposed scheme to take consideration about the two objectives: the image quality and the file size expansion. The strategy of multi-objective optimization is carried based on other schemes, it selects the optimized combination of signals according to different division of cover in different schemes and then the additional data is embedded into the selected signals. The signal of this paper is sized 8${\times}$8 DCT coefficients, and it can also be other sizes. Higher image quality and lower file size expansion under a given payload constraint can be achieved by using the multi-objective optimization strategy to choose the combination of signals for embedding. Experimental results show that the proposed method can yield better performance than some state-of-the-art methods \cite{IEEEexample:huang2016reversible} and \cite{IEEEexample:hou2018reversible}. What's more, the cover can be any format data of the JPEG image, such as the coding domain, in the future, we will make improvement on the coding domain combined the multi-objective optimization.

\section*{Acknowledgment}
\par This research work is partly supported by National Natural Science Foundation of China (61872003, U1636206, 61860206004).

\bibliographystyle{IEEEtran}
\bibliography{IEEEabrv,IEEEexample}

%
\begin{IEEEbiography}[{\includegraphics[width=1in,height=1.25in,clip,keepaspectratio]{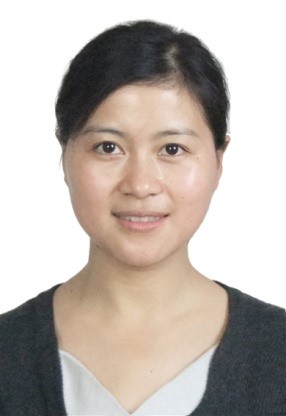}}]{Zhaoxia Yin}
received her B.Sc., M.E. and Ph.D. from Anhui University in 2005, 2010 and 2014 respectively. She is an IEEE/ACM/CCF member, CSIG senior member and the Associate Chair of the academic committee of CCF YOCSEF Hefei 2016-2017. She is currently working as an Associate Professor and a Doctoral Tutor in School of Computer Science and Technology at Anhui University. She is also the Principal Investigator of two NSFC Projects. Her primary research focus including Information Hiding, Multimedia Security and she has published many SCI/EI indexed papers in journals, edited books and refereed conferences.
\end{IEEEbiography}
\begin{IEEEbiography}[{\includegraphics[width=1in,height=1.25in,clip,keepaspectratio]{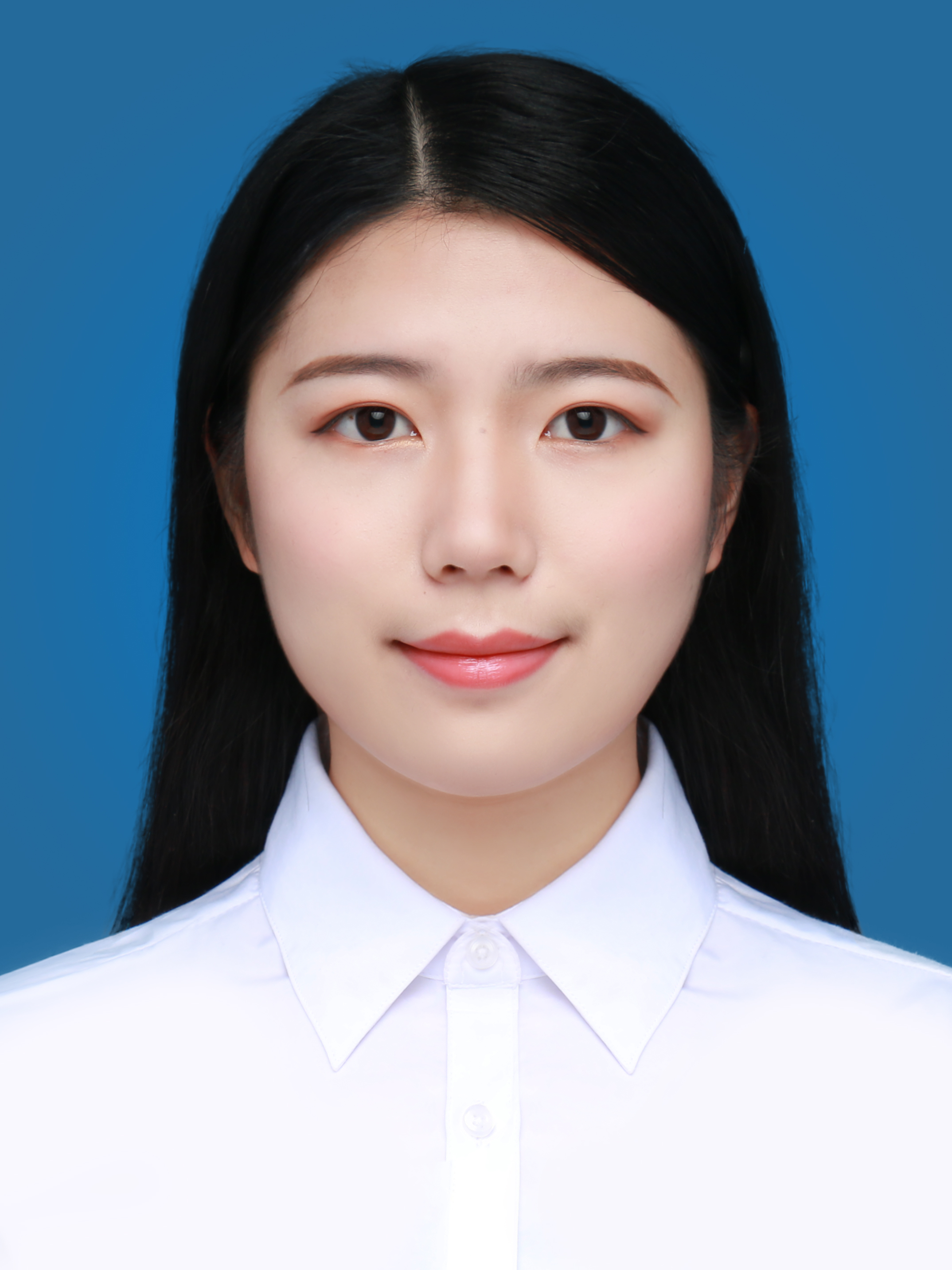}}]{Yuan Ji}
received her bachelor degree in the School of Computer Science and Technology in 2017 and now is a master student in the School of Computer Science and Technology, Anhui University. Her current research interests include reversible data hiding.
\end{IEEEbiography}
\vspace{-35em}
\begin{IEEEbiography}[{\includegraphics[width=1in,height=1.25in,clip,keepaspectratio]{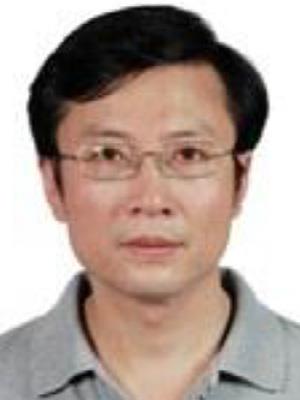}}]{Bin Luo}
received his BEng. and MEng. degrees in electronics from Anhui university, China. In 2002, he was awarded the PhD degree in Computer Science from the University of York, UK. He is currently a full professor at Anhui University. He is the chair of IEEE Hefei Subsection, and an associate chair of IAPR TC15. He serves as the editor-in-chief of the Journal of Anhui University (Natural Science Edition), an associate editor of several international journals, including Pattern Recognition, Pattern Recognition Letters, Cognitive Computation and International Journal of Automation and Computing. He was the guest editors for the Journal Special Issue of the Pattern Recognition Letters and Cognitive Computation. His current research interests include pattern recognition and digital image processing. In particular, he is interested in structural pattern recognition, graph spectral analysis, image and graph matching. He has published about 500 research papers in journals, edited books and refereed conferences. Some of his papers were published in the journals of IEEE TPAMI, IEEE TIP, Pattern Recognition, Pattern Recognition Letters and Neurocomputing, and the conferences of CVPR, NIPS, IJCAI and AAAI conferences.
\end{IEEEbiography}

\end{document}